\documentclass[a4paper]{jpconf}
\usepackage{graphicx}
\usepackage{physics}
\usepackage{bbold}
\usepackage{amsmath}
\usepackage{mathtools}
\usepackage{caption}
\usepackage{subcaption}
\usepackage{booktabs}
\usepackage{tabularray}
\usepackage[dvipsnames]{xcolor}

\usepackage[
backend=biber,
style=nature,
sorting = none
]{biblatex}
\begin{document}

\title{Beam shifts and eigenpolarisations for the reflection of vortex beams from homogeneous magnetic surfaces}
\author{Mairi Gilmour$^1$, Sarah Croke$^1$, Jörg B.~Götte$^{1,2,3}$.}
\address{$^{1}$ School of Physics and Astronomy, University of Glasgow, Glasgow, G12 8QQ, UK.}
\address{$^{2}$ Max Planck Institute for the Physics of Complex Systems, Nöthnitzer Straße 38, 01187 Dresden, Germany.}
\address{$^{3}$ Institut für Festkörperphysik, Leibniz Universität Hannover, Appelstraße 2, 50167 Hannover, Germany.}
\ead{m.gilmour.2@research.gla.ac.uk}
\date{\today}
\begin{abstract}

Recent experiments with extreme ultraviolet vortex beams reflecting from magnetic surfaces have demonstrated that vortex beams, associated with optical angular momentum, are sensitive to surface magnetisation.  Here we present a systematic treatment of vortex beam reflection from homogeneous magnetic surfaces, with analytical expressions for the Goos-Hänchen and Imbert-Fedorov beam shifts as Artmann operators. We identify that the magnetisation modulates the spin-orbit coupling, and introduces the possibility of an exceptional point, or non-hermitian degeneracy for the plane wave eigenpolarisations. We present analytical solutions for the eigenpolarisations and the shift of the vortex positions which extends the framework of singularimetry to magnetic surfaces, and detail the influence of each component of magnetisation on the beam shifts, drawing comparison to the behaviour for dielectrics.

% We present a systematic treatment of unit-order vortex beam reflection from homogeneous magnetic surfaces. Motivated by recent experiments with extreme ultraviolet vortex beams reflecting from magnetic surfaces, demonstrating that vortex beams are sensitive to surface magnetisation, we derive analytical expressions for the Goos-Hänchen and Imbert-Fedorov beam shifts as Artmann operators. We identify that the magnetisation modulates the spin-orbit coupling, and introduces the possibility of an exceptional point (EP), or non-hermitian degeneracy for the plane wave eigenpolarisations. We present analytical solutions for the eigenpolarisations, and detail the influence of each component of magnetisation on the beam shifts, drawing comparison to the behaviour for dielectrics. 

\end{abstract}

\section{Introduction}
\label{sec:introduction}

    The recent observation of magnetic helicoidal dichroism (MHD) using extreme ultraviolet (XUV) vortex beams \cite{fanciulli_observation_2022, fanciulli_magnetic_2025} has opened new avenues for probing magnetic structures through the vorticity of light \cite{allen_orbital_1992}. In these experiments, the differential absorption of vortex beams with opposite handedness encodes information about the magnetisation configuration of non-homogeneous surfaces. This discovery follows a broader trend of using structured light \cite{babiker_chapter_2021, fang_ultrafast_2025} to probe surface and material properties with enhanced sensitivity \cite{cheng_metrology_2025}. However, interpreting such experiments requires understanding not just the dichroic absorption, but also how the reflection of vortex beams from magnetized surfaces differs from the well-studied dielectric case. Here we explore the reflection of unit order scalar vortex beams, carrying orbital angular momentum, from spatially homogeneous magnetic surfaces, where throughout we refer to scalar beams as having spatially homogeneous polarisation structure \cite{babiker_chapter_2021}. 

    Magnetic surfaces display different reflection characteristics when compared to dielectric reflection, in that the magneto-optical Kerr effect (MOKE) \cite{sato_fundamentals_2022}, the reflection equivalent of Faraday rotation for transmission, allows for polarisation mixing between the two dielectric plane wave eigenpolarisations $p$ and $s$ polarisation. 
    %The MOKE is currently utilised in the imaging of magnetization on ultra-thin films, magnetic domain imaging, spintronics applications, and time-resolved magnetic imaging~\cite{sato_fundamentals_2022, fanciulli_magnetic_2025}.
    The optical angular momentum (OAM) of vortex beams possessing helical wavefronts and a phase singularity, or phase vortex, are where every photon in the beam carries an OAM of $l \hbar$.
    To analyse these magnetisation dependent effects, optical vortex beams carrying orbital angular momentum $l$ provide a particularly powerful probe. 

    The polarization mixing behaviour has profound consequences for optical beam shifts: diffractive corrections to the laws of geometric optics that arise when a light beam of finite extent is considered instead of a plane wave~\cite{bliokh_gooshanchen_2013}. These shifts are broken into the Goos-Hänchen (GH) shift parallel to the plane of incidence and Imbert-Fedorov (IF) shift transverse to the plane of incidence~\cite{bliokh_gooshanchen_2013}. 
    It is possible to identify two distinct operators (introduced as Artmann operators in \cite{gotte_generalized_2012, dennis_analogy_2012}) responsible for the GH and IF shifts. For dielectric surfaces a comprehensive analytic framework has been developed for these which links the positions of unit-charge ($\ell = \pm 1$) vortices after reflection directly to both the GH and IF shifts~\cite{dennis_topological_2012, dennis_analogy_2012}. This technique, called singularimetry, enables direct retrieval of surface characteristics including refractive index, tilt, and higher-order optical aberrations from measured vortex positions~\cite{dennis_topological_2012}, and has been recently confirmed experimentally~\cite{barros_observation_2024} for dielectric surfaces.
    
    As we shall show, these Artmann operators also encode the eigenpolarizations for a vortex beam reflecting from the surface. These eigenpolarisations are the set of polarizations for which the spatial structure is preserved upon reflection. For magnetic surfaces, the magnetisation induced polarization mixing fundamentally alters the eigenpolarisations selected by the surface. %These eigenpolarisations are useful in finding the complete reflection behaviour of the surface. including determining the geometric phase, the additional phase acquired upon reflection due to change in polarisation state \cite{gutierrez-vega_pancharatnamberry_2011}. 
    We find magnetisation has consequences for the non-Hermitian behaviour of reflection, and propose how singularimetry could be utilised to probe non-Hermitian physics utilising reflection from magnetised surfaces \cite{modak_inhomogeneous_2024, el-ganainy_non-hermitian_2018, gotte_eigenpolarizations_2014}. 
    
    %However, for magnetic surfaces, the magnetisation induced polarization mixing fundamentally alters this picture. The eigenpolarizations are no longer necessarily linear and depend on the magnetization direction, while the reflection matrix acquires off-diagonal elements that couple $s$ and $p$ components. This mixing modifies how beam shifts manifest and introduces magnetization-dependent contributions absent in the purely dielectric case.

    In this work, we extend the framework to homogeneous magnetic surfaces within the linear MOKE regime. 
    %By performing a first-order Taylor expansion of the magneto-optical reflection matrix, 
    We derive analytical expressions for the Artmann operators of the magnetic reflection matrix, which account for magnetization-dependent polarization mixing. We show that the GH and IF shifts exhibit distinct magnetization-dependent features absent in dielectric reflection.
    %: transverse magnetization introduces an additional shift resonance for beams reflecting from optically denser magnetic media, while polar magnetization produces non-zero shifts at normal incidence due to polarization rotation effects. For longitudinal and polar magnetization, which enable $s$-$p$ polarization mixing, we find that post-selection on orthogonal polarizations yields significantly enhanced shifts compared to the dielectric case, though at reduced intensity. 
    Furthermore, we determine analytical expressions for the eigenpolarizations of both plane waves and unit-charge vortex beams reflecting from homogeneous magnetic surfaces. These eigenpolarizations illuminate the properties of magnetic surfaces as reflecting elements 
    %and enable complete characterization of the total phase change—dynamical plus geometric—acquired upon reflection \cite{gutierrez-vega_pancharatnamberry_2011} 
    as well as linking the magnetisation behaviour with a modified non-Hermitian character.

    The paper is organized as follows: 
    Section~\ref{sec:beamreflection} covers the background of beam reflection and dielectric singularimetry. Section ~\ref{sec:eigpoltheory} defines how we find the eigenpolarisations of plane waves and unit-order vortex beams, and defines the vortex centroid shift. Section~\ref{sec:singularimetry} develops the singularimetry formalism for homogeneous magnetic surfaces, deriving the shift matrices and analyzing magnetization-dependent vortex centroid behaviour for different magnetization configurations. Section~\ref{sec:eigenpolarizations} determines eigenpolarizations for the plane wave case, and details the additional non-Hermitian character found for reflection from homogeneous magnetic surfaces. Section~\ref{sec:vortex_eigenpolarizations} determines the eigenpolarisations of the unit order vortex beam, comparing magnetic and dielectric cases and again the non-Hermitian character. Section~\ref{sec:discussion} discusses implications for magnetic surface characterization and potential extensions to non-homogeneous magnetization patterns.

\section{Beam reflection and singularimetry for dielectric surfaces}
\label{sec:beamreflection}

    Optical vortex beams are characterized by a helical phase structure $\mathrm{e}^{\mathrm{i}\ell\phi}$, where the integer topological charge $\ell$ determines the orbital angular momentum of $\ell\hbar$ per photon~\cite{allen_orbital_1992, } and $\phi$ is the azimuthal angle around the beam axis. This phase structure produces a phase singularity, a point of undefined phase and zero intensity, on the beam axis, whose position after reflection directly encodes information about beam shifts and surface properties. 
    %Vortex singularimetry accounts for the beam's finite spatial extent by decomposing it into plane wave components with varying local incidence angles~\cite{gotte_generalized_2012}.
    %with the vortex position providing an analyser-dependent measurement when the reflected beam is post-selected onto a specific polarization~\cite{gotte_generalized_2012}.  

    The positions of the optical vortices after reflection can be utilised to measure surface properties such as refractive index or aberration terms. An analytical approximation of these vortex positions after reflection from a flat dielectric surface was developed in \cite{dennis_topological_2012}. The consideration that real vortex beams are composed of a bundle of plane waves and hence physically localised was a key consideration.
    
\begin{figure}[t]
    \centering
    \includegraphics[width= 0.7\textwidth]{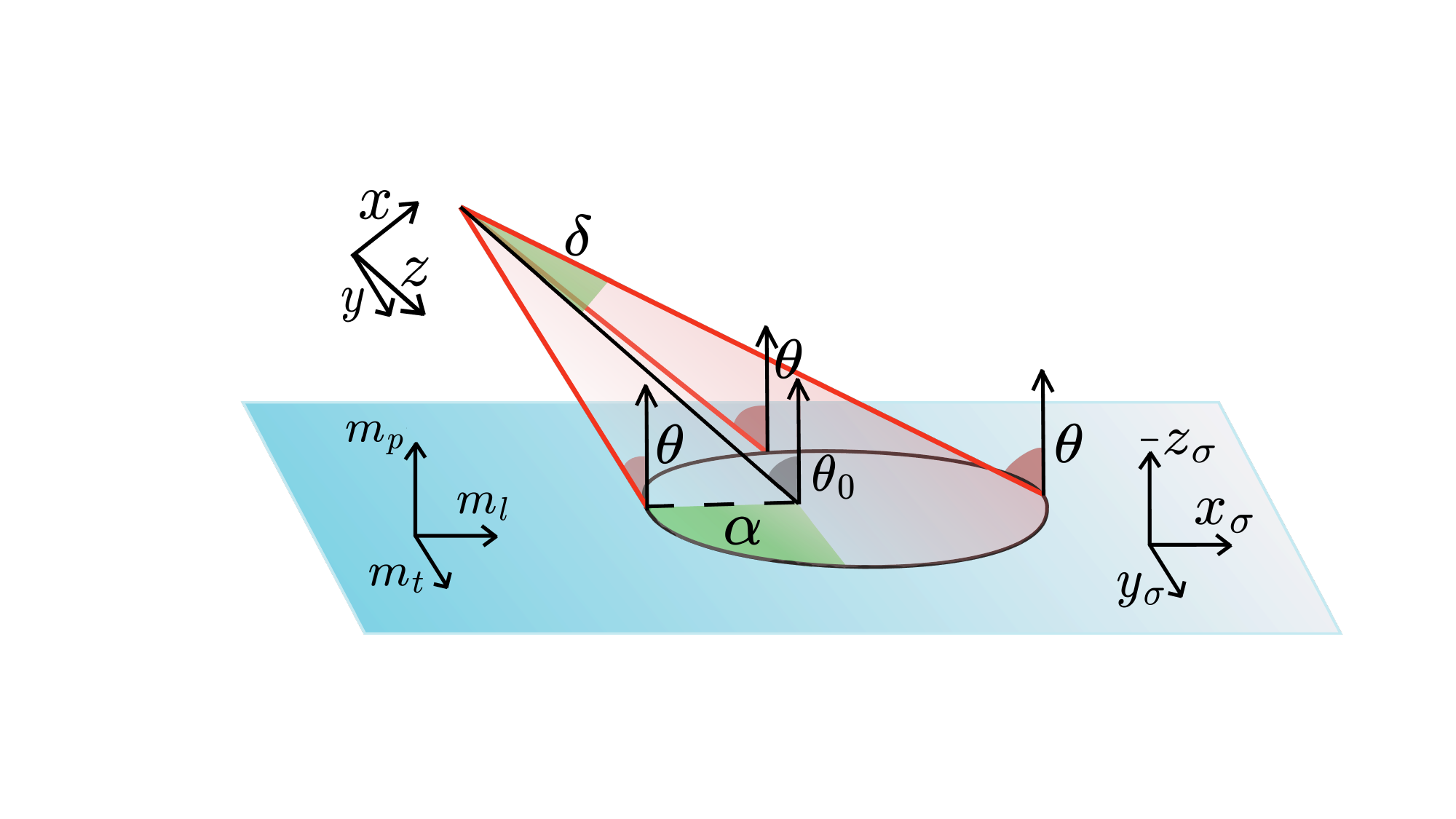}
    \caption{Diagram outlining the beam geometry, depicting the red cone of wave-vectors for an incident vortex beam. The Fourier variable $\delta$, denotes the opening angle of the conical spread of the cone of wave-vectors $k(\theta_0,\alpha,\delta)$, where $\alpha$ is the radial angle. 
    %Taking $\delta=0$ recovers the delta-function `cone' of an incident plane wave, with a single wave-vector. 
    The incidence angle of the central wave vector is given by $\theta_0$, and the plane of incidence this defines in blue highlights how the local angle of incidence $\theta(\theta_0,\alpha,\delta)$ varies around the beam cone. %changes as a result of the beam being made of a bundle of plane waves. 
    The directions of the magnetisation components are also shown relative to the central plane of incidence.
    %with the $x_\sigma$-direction along $m_l$ and the $y_\sigma$-direction along $m_t$. 
    The two coordinate systems are highlighted: the intrinsic beam coordinate system, $x,y$ and $z$, and the surface lab coordinate system $x_\sigma, y_\sigma$ and $z_\sigma$ \cite{fanciulli_electromagnetic_2021}. Conversion between the two is given in the supplemental material.}
    \label{Beam Geometry Figure}
    \end{figure}

    %Singularimetry is an analytical treatment for the reflection of optical vortex beams which calculates an approximation of the reflected vortex position or positions, which would normally be measured through interferometric techniques, for a given dielectric surface where the angle of incidence and surface properties are known \cite{dennis_topological_2012}. 
    %In this paper, we shall extend this treatment to homogeneous magnetic surfaces, where first we recount the geometry necessary to describe the beam's finite spatial structure, and then the differing reflection matrix adopted for reflection from homogeneous magnetic surfaces. 
    To account for the physical extent of the beam, it can be decomposed into plane wave components with varying local incidence angles, where we parametrise the local angle of incidence as in~\cite{dennis_topological_2012}:
    \begin{equation}\label{local theta}
      \theta(\theta_0,\alpha,\delta)=\cos^{-1}(\cos\delta\cos\theta_0-\cos\alpha\sin\delta\sin\theta_0),
    \end{equation}
    where $\theta_0$ is the central angle of incidence, $\delta\ll 1$ the spectral spread of the paraxial beam, and $\alpha$ the azimuthal angle around the beam cone (Figure~\ref{Beam Geometry Figure}). 

    Throughout our analysis, we adopt the Jones matrix calculus adapted from~\cite{fanciulli_electromagnetic_2021}, which models how a plane wave reflects from a magnetised surface via a $2\times 2$ reflection matrix $\mathbf{R}$ in the $sp$ polarisation basis. For a finite beam, the angle-dependence of $\mathbf{R}$ must be made explicit by accounting for the local angle of incidence of each plane-wave constituent of the beam spectrum. The reflection matrix for a dielectric surface is then

    \begin{equation}\label{diR0}
    \mathbf{R}\big|_{\delta=0} = 
    \begin{pmatrix}
    r_p  & 0 \\
    0 & r_s
    \end{pmatrix},{}
    \end{equation}
    where $r_p$ and $r_s$ are the dielectric Fresnel reflection coefficients \cite{born_principles_2019}. Eq. \eqref{diR0} is the plane-wave ($\delta=0$) reflection matrix, later referred to as $ \mathbf{R}_0$.

    Because each plane-wave component of the beam arrives at a slightly different local incidence angle, the polarisation unit vectors likewise vary around the angular spectrum of the beam. Defining the surface normal $\mathbf{n}_s=(-\sin\theta_0,0,\cos\theta_0)$ and the local unit wavevector $\boldsymbol{\kappa}=(\sin\delta\cos\alpha,\sin\delta\sin\alpha,\cos\delta)$, the local $s$ and $p$ unit vectors follow from the standard construction~\cite{gotte_generalized_2012}:
    \begin{equation}\label{sp unit vectors}
      \mathbf{e}_s = \frac{\mathbf{n}_s\times\boldsymbol{\kappa}}{|\mathbf{n}_s\times\boldsymbol{\kappa}|}, \qquad \mathbf{e}_p = \mathbf{e}_s\times\boldsymbol{\kappa}.
    \end{equation}
    The corresponding diagonal projectors $P_{s}=\mathbf{e}_s\,\otimes\,\mathbf{e}_s$ and $P_{p}=\mathbf{e}_p\,\otimes\,\mathbf{e}_p$ suffice for a purely dielectric surface. 

    \begin{equation}\label{dielectric R}
    \begin{gathered}
      \mathbf{R} = r_s\, \mathbf{P}_s + r_p\, \mathbf{P}_p.
    \end{gathered}
    \end{equation}
    
     For dielectric surfaces, where the reflection matrix contains only the diagonal Fresnel coefficients $r_p$ and $r_s$ with no coupling between polarization components, $s$ and $p$ polarized light remains unchanged in polarization upon reflection. These linear polarizations are therefore the eigenpolarizations of dielectric reflection, and beam shifts, such as the Goos-Hänchen and Imbert-Fedorov effects \cite{gotte_generalized_2012}, can be understood in terms of these two independent reflection channels.

    The Goos-Hänchen (GH) beam shift parallel to the plane of incidence arises because the different plane waves in an angular decomposition of the beam are reflected slightly differently as both the phase and magnitude of $r_p$ and $r_s$ depend on the angle of incidence. The Imbert-Fedorov (IF) shift transverse to the plane of incidence affects any beams which are not purely $s$ or $p$ polarized. Its origins lie in the linking between spatial and polarization degree of freedom due to the transversality conditions, which causes left and right circular polarization components to be displaced with respect to one another \cite{bliokh_conservation_2006}. 

     An analytic approximation of the reflected beam field is then constructed by Taylor-expanding the angle-dependent reflection matrix $\mathbf{R}(\delta,\alpha)$ in the small beam-spread parameter $\delta$. Each power of $\delta$ carries a definite $\alpha$-symmetry, and the leading first-order terms define the two Artmann operators responsible for the Goos-H\"{a}nchen shift in the $x$-direction and Imbert-Fedorov shift in the $y$-direction \cite{dennis_topological_2012, gotte_generalized_2012}.

    The angular symmetry of the $\alpha$-independent Fresnel coefficients,  given the $\alpha$ dependence comes from the projectors only, ensures that, to first order, $\mathbf{R}$ decomposes as~\cite{gotte_generalized_2012, dennis_analogy_2012},
    \begin{equation}\label{Artmann def}
      \mathbf{R}(\delta,\alpha) = \mathbf{R}_0 + \delta\cos\alpha\,\mathbf{R}_\text{GH} + \delta\sin\alpha\,\mathbf{R}_\text{IF} + \mathcal{O}(\delta^2),
    \end{equation}
    where henceforth we suppress denoting the explicit evaluation at $\delta=0$. The operator $\mathbf{R}_\text{GH}$ captures the $\cos\alpha$ part of $\partial \mathbf{R}/\partial\delta$ and generates the Goos-H\"{a}nchen shift; $\mathbf{R}_\text{IF}$ captures the $\sin\alpha$ part and generates the Imbert-Fedorov shift. This can be understood from the relation $\mathrm{e}^{\mathrm{i}\alpha}=\cos{\alpha}+\mathrm{i}\sin{\alpha}=x+\mathrm{i}y$ where the shift in the $x$ direction, associated with $\cos{\alpha}$ can be related to the GH shift and similarly for the IF shift. Their physical interpretation is immediate for dielectric surfaces: $\mathbf{R}_\text{GH}$ involves derivatives of the Fresnel coefficients with respect to angle of incidence and reflects the angular dispersion of the reflection phase, while $\mathbf{R}_\text{IF}$ involves only the difference $(r_p-r_s)$ and reflects the transversality-enforced spin-orbit coupling~\cite{bliokh_gooshanchen_2013}. For dielectric surfaces their exact form is \cite{gotte_generalized_2012} 

    \begin{equation}\label{GH IF shifts dielectric}
    \begin{gathered}
      \mathbf{R} = \underbrace{\begin{pmatrix}
    r_p  & 0 \\
    0 & r_s
    \end{pmatrix}}_{\mathbf{R}_0}
    +\delta\cos\alpha\underbrace{\begin{pmatrix}
    r_p' & 0   \\
    0  & r_s'
    \end{pmatrix}}_{\mathbf{R}_\text{GH}}
    +
    \delta\sin\alpha\underbrace{\begin{pmatrix}
    0& (r_p-r_s)\cot\theta_0 \\
    (r_p-r_s )\cot\theta_0 & 0
    \end{pmatrix}}_{\mathbf{R}_\text{IF}},
    \end{gathered}
    \end{equation}

    where we define $r_{p,s}'=\partial \, r_{p,s} / \partial \,\theta_0 $. Singularimetry allows for an analytic approximation of these shifts of the intensity centroid through relation to the shift of a reflected vortex for an $l=\pm1$ beam \cite{dennis_topological_2012}. To extract the vortex centroid shift, we now apply this framework to a unit-charge vortex beam. The input unit vortex beam, $\psi(x,y,z)_I$, where the phase vortex contained in the $\mathrm{e}^{\pm \mathrm{i} \phi}$ term with $l=\pm 1$ is associated with a positive and negative orbital angular momentum respectively \cite{allen_orbital_1992}. We assume a narrow spectrum in $k$-space \cite{shen_optical_2019,fanciulli_electromagnetic_2021}, $\delta\ll1$, is given in Eq.~\ref{input psi} for $l=+1$:

    \begin{equation}\label{input psi}
    \begin{gathered}
      \psi(x,y,z)_I \,=  A(x,y,z) \mathrm{e}^{\mathrm{i}(\omega t-kz) \pm \mathrm{i} \phi} \, \mathbf{E}.
    \end{gathered}
    \end{equation}

    The polarization vector $\mathbf{E} = (E_p, E_s)$ with respective $s$ and $p$ polarisation components determines the polarisation of the light before reflection, $A(x,y,z)$ the spatially varying amplitude factor, $\phi$ the real space azimuthal angle, and $\mathrm{e}^{\mathrm{i}(\omega t-kz)}$ the beam free-space propagation term. 

    To calculate the reflected beam we consider the angular spectrum or Fourier decomposition at $z = 0$, at the surface, and $t = 0$, effectively dropping the spatial and temporal propagation terms and only keeping the phase vortex term, its polarisation vector and an $\alpha$-independent, rotationally symmetric spectrum term $\sigma(\delta)$ which defines the beam spread in k-space to give an input beam in Fourier space of the form for $l=+1$:
    
    \begin{equation}\label{input psi polar}
    \begin{gathered}
      \psi(\alpha, \delta)_I \,=  \sigma(\delta) \mathrm{e}^{+ \mathrm{i} \alpha}  \, \mathbf{E}.
    \end{gathered}
    \end{equation}
    We now take the inverse Fourier transform in terms of the Fourier polar variables $\delta$ and $\alpha$ of $\mathbf{R} \, \psi(\alpha, \delta)_I $ to give the reflected field $\psi(x,y)_R$ at a point $(x,y)$ at the interface:

    \begin{equation}\label{reflected dielectric field}
    \begin{aligned}
      \psi(x,y)_R \,&= \int_0^{\pi/2}\int^\pi_{-\pi} \sigma(\delta) \delta \mathrm{e}^{\mathrm{i}k\sin{\delta} (x\cos{\alpha}+ y\sin{\alpha}) -\mathrm{i}\alpha} \,\mathbf{R}\mathbf{E} \,d\delta d\alpha,  \\
      &=\int_0^{\pi/2}\int^\pi_{-\pi} \sigma(\delta) \delta \mathrm{e}^{\mathrm{i}kr\delta\cos(\alpha-\phi) -\mathrm{i}\alpha} \,\mathbf{R}\mathbf{E} \,d\delta d\alpha, 
      \\
      \text{where} \quad \,&x=r\cos\phi, \qquad y=r\sin\phi, \qquad r=\sqrt{x^2+y^2},
    \end{aligned}
    \end{equation}

where it is identified as in \cite{dennis_topological_2012} that $\sin \delta=\delta$ for $\delta\ll 1$, the additional factor of $\delta$ comes from the Jacobian of transforming into polar coordinates \cite{barros_observation_2024} and the sign of $\mathrm{e}^{\mathrm{i}\alpha}$ flips upon reflection to preserve that the beam is still propagating in the positive $z$ direction.

$\mathbf{R}\mathbf{E}$ given in Eq.~\eqref{reflected dielectric field} generates the full reflected field, where the object $[\mathbf{R}\mathbf{E}](\alpha,\delta)$  is the full $k$ dependent $2\,\text{x}\,1$ vector of reflected polarisations for beam, the same shape as the full vector input $\mathbf{E}$ itself. Integrating this object over $\alpha$ and $\delta$ recovers the full field with its spatially varying structure, $\psi(x,y)_R $.

For the purposes of scalar singularimetry, it is necessary to consider instead the reflected field scalar  $\mathbf{F^\text{*}}\mathbf{R}\mathbf{E}$, where the reflected beam is post-selected by an analyser $\mathbf{F} = (F_p, F_s)$ \cite{dennis_analogy_2012}. As the vortex position and hence centroid shift are $\mathbf{F}$ analyser dependent, we will now calculate the centroid shift and vortex position for the $\mathbf{F}$ analysed polarisation component of the field, $\psi(x,y)_F$. 
\begin{equation}\label{reflected dielectric scalar}
\begin{gathered}
  \psi(x,y)_F \,= \int_0^{\pi/2}\int^\pi_{-\pi} \sigma(\delta) \delta \mathrm{e}^{\mathrm{i}kr\delta\cos(\alpha-\phi) -\mathrm{i}l\alpha} \mathbf{F^\text{*}}\mathbf{R}\mathbf{E} \,d\delta d\alpha,
\end{gathered}
\end{equation}
In this case, the vortex centroid of the $\psi(x,y)_F$ component of the field, can be directly accessed as a single point \cite{gotte_generalized_2012}.

\section{Calculating eigenpolarisations for plane waves and unit-order scalar vortex beams}
\label{sec:eigpoltheory}

Now we have recovered the full reflected field, we can determine the eigenpolarisations of the surface. 
These describe the set of polarisations for which the vectorial direction remains unchanged upon reflection from the surface \cite{garza-soto_geometric-phase_2020}. As we consider only the set of transverse polarisations, there are two transverse eigenpolarisations for the surface. 
%, with the amplitude and overall total phase changes described by the corresponding eigenvalues and eigenpolarisations \cite{garza-soto_geometric-phase_2020}.  

%%% 
The eigenpolarisations of a unit-charge vortex beam emerge naturally from the first-order approximation to the full reflected field. Substituting Eq.~\eqref{Artmann def} into Eq.~\eqref{reflected dielectric field} and evaluating the $\alpha$-integrals using the Bessel generating-function identity (with details in the supplementary), then retaining only the leading-order contributions in $\delta$ ($J(kr\delta)_0\approx 1$, $J(kr\delta)_1\approx kr\delta/2$, $J(kr\delta)_2\approx \mathcal{O}(\delta^2)$), yields for $l=+1$

\begin{equation}\label{eigpol_original}
\psi(x,y)_R \,\propto \left[ \mathrm{i} \eta\, \mathbf{R}_0 + (\mathbf{R}_\text{GH}-\mathrm{i} \mathbf{R}_\text{IF}) \right] \mathbf{E},
\end{equation}
%%%
%%% explain eta??

with $(\mathbf{R}_\text{GH}+\mathrm{i} \mathbf{R}_\text{IF})$ for $l=-1$. We define $\eta^a=(k r)^{\abs{a}}\mathrm{e}^{-\mathrm{i}a\phi}= k^{a}(x-\mathrm{i}y)^{a}$ for $a\ge 0$ and $(\eta^*)^{a}=(k r)^{\abs{a}}\mathrm{e}^{-\mathrm{i}a\phi}= k^{a}(x+\mathrm{i}y)^{a}$ for $a<0$. $(\eta^{*})^a$ is the $a$ power of the complex conjugate of $\eta $, which is the complex transverse real-space coordinate. 
It is possible to separate out the plane wave reflection matrix from Eq. \eqref{eigpol_original} so it acts on $\mathbf{E}$ first: 

\begin{equation}\label{eigpol_seperate}
\begin{aligned}
\psi(x,y)_R \,&\propto \left[ \mathrm{i} \eta\, \mathbf{I} + (\overline{\mathbf{R}}_\text{GH}\mp\mathrm{i}\overline{\mathbf{R}}_\text{IF}) \right] \mathbf{R}_0\mathbf{E},\\
&= \left[ \mathrm{i}\eta\, \mathbf{I} + (\overline{\mathbf{R}}_\text{GH}\mp\mathrm{i}\overline{\mathbf{R}}_\text{IF}) \right] \mathbf{E}_i,
\end{aligned}
\end{equation}

where $\overline{\mathbf{R}}_\text{GH,IF}=\mathbf{R}_\text{GH,IF} \mathbf{R}_0^{-1}$ are referred to as the Artmann operators as in \cite{gotte_generalized_2012}, it is these matrices which generate the GH and IF shifts. We identify  $\mathbf{R}_0^{-1}\mathbf{R}_0=\mathbf{I}$ where $\mathbf{R}_0^{-1}$ is the inverse of the plane wave reflection matrix $\mathbf{R}_0$ and $\mathbf{I}$ the identity, and $\mathbf{E}_i=\mathbf{R}_0\mathbf{E}$ the set of polarisations after the plane wave reflection component has acted. To find the set of eigenpolarisations for plane waves, it is then as simple as finding the eigendecomposition of $\mathbf{R}_0\mathbf{E}=\lambda\mathbf{E}$ where $\lambda$ are then the eigenvalues when $E\in (\nu_1,\nu_2)$ where $\nu_{1,2}$ are the set of transverse eigenpolarisations. For the eigenpolarisations of the unit-order vortex beam, we define these as the set of polarisations following the eigendecomposition for $\mathbf{E}_i$ in Eq. \eqref{eigpol_seperate}, as such factoring out the plane wave reflection. For decomposing into separate GH or IF shifts, find the eigenpolarisations for $\overline{\mathbf{R}}_\text{GH,IF}\mathbf{E}_i$.
 
For a homogeneous input $l=\pm 1$ vortex beam polarisations $\mathbf{E}_i$ in general, the action of Eq. \eqref{eigpol_seperate} produces a spatially varying, inhomogeneous reflected polarisation. This is because the $p$ and $s$ components generally have differing spatial dependence and hence produce an inhomogeneous reflected polarisation. For the case of input eigenpolarisations for $l=\pm 1$ vortex beams, it is found that this spatial dependence for the $p$ and $s$ components is identical up to a multiplicative factor. 

As a result, the spatial dependence can be factored out, and we recover a set of homogeneous polarisations after reflection. These are not necessarily in the same vectorial direction as what would be experimentally incident on the surface $\mathbf{E}$, but are always in the same vectorial direction as $\mathbf{E}_i$ and hence the actual experimental input polarisation $\mathbf{E}$ can always be recovered. We detail these vortex beam eigenpolarisations, along with the plane wave counterparts recovered from the eigendecomposition of $\mathbf{R}_0\mathbf{E}$ for both dielectric and homogeneous magnetic surfaces in Section~\ref{sec:eigenpolarizations}. 

The scalar equivalent to Eq. \eqref{eigpol_original} can be expressed as 
%For the rest of this section, we will deal with the reflected field scalar $F^*RE$, useful for calculating both the vortex centroid and the centre of intensity shifts of a post-selected component \cite{gotte_generalized_2012}, with the understanding that the full field $RE$ can be calculated using the same method just in a component-wise manner. 

%We now with to utilise the process of singularimetry for homogeneous magnetic surfaces, to produce an analytic approximation of the field after reflection, and hence the integral in eq. \ref{reflected dielectric scalar}. This involves Taylor expanding the reflected field scalar $F^*RE$ up to order $\abs{l}$ in the Fourier variable $\delta$, from which we can exploit the symmetries of the beam and separately deal with the $\alpha$ integration to produce a polynomial of order $\abs{l}$, the roots of which give the position (or positions) of the vortex after reflection given the vortex is a point of zero intensity \cite{gotte_generalized_2012}. 

\begin{equation}\label{reflected_scalar_matrixform}
\begin{gathered}
\psi(x,y)_F \,\propto \mathbf{F^\text{*}}\left[ \mathrm{i} \eta\, \mathbf{R}_0 + (\mathbf{R}_\text{GH}\mp\mathrm{i} \mathbf{R}_\text{IF}) \right] \mathbf{E}
= \mathrm{i} \eta\, R_0^{EF} + (R_\text{GH}^{EF}\mp\mathrm{i}R_\text{IF}^{EF})
,
\end{gathered}
\end{equation}

where $R_\text{0,GH,IF}^{EF}= \mathbf{F^\text{*}}\, \mathbf{R}_\text{0,GH,IF}\,\mathbf{E}$ is the $\mathbf{F}$ analysed polarisation component of the reflected field generated by reflection matrix $\mathbf{R}_\text{0,GH,IF}$ with input polarisation $\mathbf{E}$. The objects $R_\text{0,GH,IF}^{EF}$ are scalars, where solving for $\eta$ where Eq. \eqref{reflected_scalar_matrixform} $=0$ returns the position of the reflected vortex, as the vortices are exactly located at the intensity zeros. It should be noted here that this scalar form of Eq. \eqref{reflected_scalar_matrixform} is now the same form as separately derived by Barros \textit{et al} in \cite{barros_observation_2024}. The vortex shift can be directly related to the centroid shift for the $\mathbf{F}$ analysed polarisation component as in \cite{gotte_generalized_2012}, where the $x$-direction component gives the Goos-H\"{a}nchen shift and the $y$-direction component the Imbert-Fedorov shift. From the corresponding Artmann operators, the centroid shift can be evaluated for both dielectric and spatially homogeneous magnetic surfaces, with these results detailed in Section \ref{sec:singularimetry}.

\section{Beam shifts for reflection from a homogeneous magnetic surface}
\label{sec:singularimetry}
     Reflection from magnetic surfaces must account for polarization rotation due to surface magnetism through the magneto-optical Kerr effect (MOKE)~\cite{sato_fundamentals_2022}. The effect decomposes the surface magnetization vector into three orthogonal components relative to the surface (see Figure~\ref{Beam Geometry Figure}): transverse magnetization, which acts only on light polarized parallel to the plane of incidence ($p$ polarized); and longitudinal and polar magnetization components, which enable polarization mixing between the transverse polarization components parallel and orthogonal to the plane of incidence ($p$ and $s$ polarized respectively). 

    The three MOKE components enter the reflection matrix through the Fresnel coefficients $r_0^t$ (transverse), $r_{ps}^l$ (longitudinal), and $r_{ps}^p$ (polar), for a surface with relative refractive index $n$ \cite{ oppeneer_chapter_2001}:
    \begin{equation}\label{all fresnel}
    \begin{aligned}
    %r_s & =  \frac{\cos{\theta}-\sqrt{n^2-\sin^2{\theta}} }{\cos{\theta}+\sqrt{n^2-\sin^2{\theta}} },\\
    %r_p & = \frac{n^2\cos{\theta}-\sqrt{n^2-\sin^2{\theta}} }{n^2\cos{\theta}+\sqrt{n^2-\sin^2{\theta}} }, \\
    r_{ps}^l & =  \frac{\mathrm{i} Q \cos{\theta}\sin{\theta} }{\sqrt{1-\frac{\sin^2{\theta}}{n^2}} \left( n\cos{\theta} + \sqrt{1-\frac{\sin^2{\theta}}{n^2}}\right) 
    \left( \cos{\theta} + n\sqrt{1-\frac{\sin^2{\theta}}{n^2}}\right)}, 
    \\ 
    r_0^t & = \frac{2\mathrm{i}Q \cos{\theta} \sin{\theta}}{n^2\cos^2{\theta} - 1 + \frac{\sin^2{\theta}}{ n^2}},
    \\
    r_{ps}^p & = 
    \frac{-\mathrm{i} Q n \cos{\theta}}
    {\left( n\cos{\theta} + \sqrt{1-\frac{\sin^2{\theta}}{n^2}}\right) 
    \left( \cos{\theta} + n\sqrt{1-\frac{\sin^2{\theta}}{n^2}}\right)}.
    \end{aligned}
    \end{equation}
    We introduce the magneto-optic constant $Q$, which is a complex valued constant of the magnetic surface and can be derived  from the off-diagonal  $\varepsilon_{xy}$ component of the dielectric permittivity tensor for magnetisation in the polar direction \cite{sato_fundamentals_2022, hunt_magnetooptic_1967,ino_terahertz_2004}:

    \begin{equation}\label{dielectric_tensor}
    \mathbf{\epsilon} = 
    \begin{pmatrix}
    \epsilon_{xx} & \epsilon_{xy} & 0 \\
    -\epsilon_{xy} & \epsilon_{xx} & 0\\
    0 & 0 & \epsilon_{zz}
    \end{pmatrix}=
    \begin{pmatrix}
    \epsilon_{xx} & -\mathrm{i} Q\epsilon_{xx} & 0 \\
    \mathrm{i} Q\epsilon_{xx} & \epsilon_{xx} & 0\\
    0 & 0 & \epsilon_{zz}
    \end{pmatrix}
    \, \,  \text{where} \, \, 
    Q= \mathrm{i} \frac{\epsilon_{xy}}{\epsilon_{xx}}.
    {}
    \end{equation}

    The Fresnel coefficients given in Eq. \eqref{all fresnel} are derived to contain terms up to linear order in $Q$ \cite{hunt_magnetooptic_1967}, and for materials where the magnetic response is much weaker than the electric, as is common in the optical regime, this constant takes a value $Q\ll 1$. The real magnetisation direction coefficients $m_t$, $m_l$, $m_p$, bounded by $m_t^2+m_l^2+m_p^2=1$, are illustrated in Figure~\ref{Beam Geometry Figure}. For spatially homogeneous magnetisation, the transformation from the lab frame to the beam frame is given in the supplemental material, and all further mention of values of $m_t, m_l$ and $m_p$ are given in the beam frame. 

    The reflection matrix for a homogenous magnetic surface is then

    \begin{equation}\label{magR0}
    \mathbf{R}\big|_{\delta=0} = 
    \begin{pmatrix}
    r_p +m_t r_0^tr_p & m_lr^l_{ps}+m_pr^p_{ps} \\
    -m_lr^l_{ps}+m_pr^p_{ps} & r_s
    \end{pmatrix},{}
    \end{equation}
    where $r_p$ and $r_s$ are the dielectric reflection coefficients \cite{born_principles_2019}. Eq.~\eqref{magR0} is the plane-wave ($\delta=0$) reflection matrix for homogeneous magnetic surfaces later referred to as $\mathbf{R}_0$. In contrast to the diagonal dielectric case, the magnetic matrix has non-zero off-diagonal elements encoding polarisation mixing~\cite{freiser_survey_1968}. We now construct the angle-dependent generalization, $\mathbf{R}(\delta,\alpha)$, required for spatially finite beams.

    The magnetic surface additionally requires the off-diagonal projectors $P_{ps}=\mathbf{e}_p\otimes\mathbf{e}_s$ and $P_{sp}=\mathbf{e}_s\otimes\mathbf{e}_p$, which map how the polarisation-mixing MOKE coefficients depend on the local beam geometry. Expanding these projectors to first order in $\delta$ and multiplying by the corresponding term in Eq.~\eqref{magR0} yields the locally dependent reflection matrix in Eq.~\eqref{magnetic R}.

    \begin{equation}\label{magnetic R}
    \begin{gathered}
      \mathbf{R} = r_s\, \mathbf{P}_s + r_p\, \mathbf{P}_p + m_t r_0^t \,\mathbf{P}_p 
      + \left( m_lr^l_{ps}+m_pr^p_{ps} \right) \mathbf{P}_{ps} + \left( -m_lr^l_{ps}+m_pr^p_{ps} \right) \mathbf{P}_{sp}.
    \end{gathered}
    \end{equation}

    For a homogeneous magnetic surface, the same expansion as in Eq.~\eqref{magnetic R} yields a similar decomposition into $\mathbf{R}_\text{GH}$ and $\mathbf{R}_\text{IF}$, which are tunable for specific magnetic metasurfaces via the $Q$ parameter in the reflection coefficients \cite{tang_realization_2018, asiri_controlling_2016, dadoenkova_controlling_2016}:
    \begin{equation}\label{GH IF shifts magnetic}
    \begin{gathered}
      \mathbf{R} = \underbrace{\begin{pmatrix}
    r_p +m_t r_0^tr_p & m_lr^l_{ps}+m_pr^p_{ps} \\
    -m_lr^l_{ps}+m_pr^p_{ps} & r_s
    \end{pmatrix}}_{\mathbf{R}_0}
    +\delta\cos\alpha\underbrace{\begin{pmatrix}
    r_p'+m_t(r^{t\prime}_0 r_p + r_0^{t}r_p' ) & m_pr^{p\prime}_{ps} +m_lr^{l\prime}_{ps}   \\
    m_pr^{p\prime}_{ps} - m_lr^{l\prime}_{ps}  & r_s'
    \end{pmatrix}}_{\mathbf{R}_\text{GH}}
    \\+
    \delta\sin\alpha\underbrace{\begin{pmatrix}
    -2  m_l r_{ps}^l\cot\theta_0 & (r_p-r_s + m_t r_0^tr_p)\cot\theta_0 \\
    (r_p-r_s + m_t r_0^tr_p)\cot\theta_0 & 2 m_p r_{ps}^p \cot\theta_0
    \end{pmatrix}}_{\mathbf{R}_\text{IF}}.
    \end{gathered}
    \end{equation}

    This is the first main result of this study: the magnetic operators inherit the same $\cos\alpha$/$\sin\alpha$ decomposition as in Eq.~\eqref{Artmann def}, and the dielectric expressions are recovered upon setting all magnetisation components to zero. These Artmann operators for spatially homogeneous magnetic surfaces display the key structural difference that magnetisation breaks the diagonal/anti-diagonal symmetry of the dielectric case: the longitudinal and polar MOKE terms populate the off-diagonals of $\mathbf{R}_\text{GH}$ (through the derivatives $m_lr_{ps}^{l\,\prime}$ and $m_pr_{ps}^{p\,\prime}$) and simultaneously introduce diagonal entries in $\mathbf{R}_\text{IF}$ (the terms $\propto m_l r_{ps}^l\cot\theta_0$ and $m_p r_{ps}^p\cot\theta_0$). The transverse magnetisation, by contrast, enters $\mathbf{R}_\text{GH}$ only through its modification of the $pp$-diagonal and $\mathbf{R}_\text{IF}$ in the off-diagonal terms adjusting the spin-orbit coupling (the term $(r_p-r_s + m_t r_0^tr_p)\cot\theta_0$). The off-diagonal population through the longitudinal magnetisation specifically introduces a non-orthogonality in the eigenvectors of even the plane wave reflection, which modifies the non-Hermitian behaviour observed for homogeneous magnetic surfaces, which is discussed in Section \ref{sec:eigenpolarizations}.

\begin{figure}[t] % t for position at the top of the current page; b for position at the bottom; p for new page
		\centering
        \begin{subfigure}[b]{0.375\linewidth} % Fig (e)
			\includegraphics[width=\linewidth]{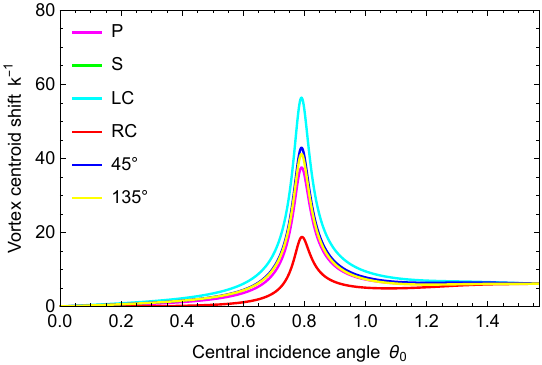}
			\caption{For a dielectric surface.}
			\label{fig2:fige}
        \end{subfigure}
        \hspace{20pt}
        \begin{subfigure}[b]{0.375\linewidth} % Fig (d)
			\includegraphics[width=\linewidth]{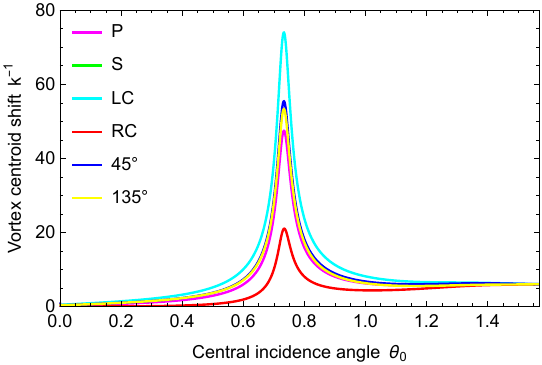}
			\caption{Magnetic surface with $m_t=1$.}
			\label{fig2:figd}
		\end{subfigure}
		   % Space between the figures
		
        \begin{subfigure}[b]{0.375\linewidth} % Fig (c)
			\includegraphics[width=\linewidth]{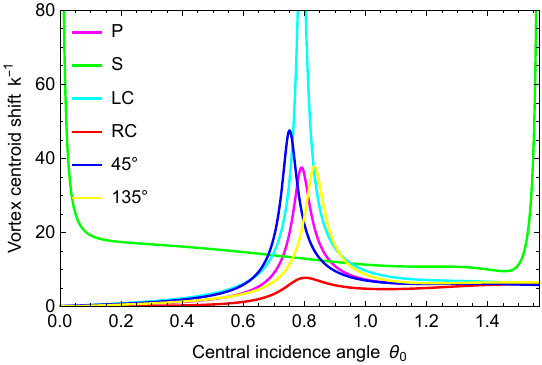}
			\caption{Magnetic surface with $m_l=1$. }
			\label{fig2:figc}
		\end{subfigure}
        \hspace{20pt}
        \begin{subfigure}[b]{0.375\linewidth} % Fig (b)
			\includegraphics[width=\linewidth]{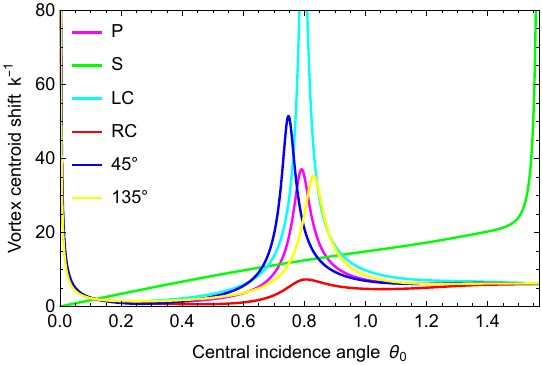}
			\caption{Magnetic surface with $m_p=1$.}
			\label{fig2:figb}
		\end{subfigure}
        
        \begin{subfigure}[b]{0.38\linewidth} % Fig (a)
			\includegraphics[width=\linewidth]{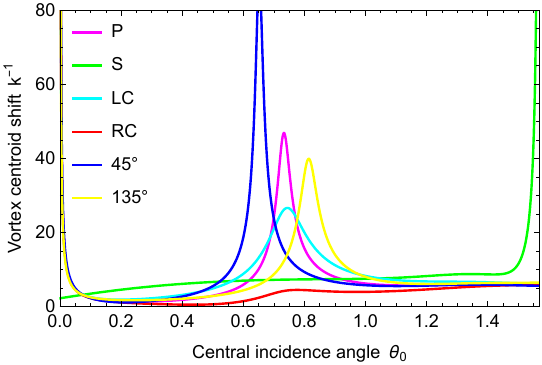}
			\caption{Magnetic surface with $m_t=m_l=m_p=1/ \sqrt{3}$.}
			\label{fig2:figa}
		\end{subfigure}
        \hspace{20pt}
        \hspace{0.38\linewidth}

		\caption{Plots of vortex centroid shift magnitude for an input $p$-polarised vortex beam, with analyser polarisations shown in the legend. a) shows the shifts for a dielectric surface with $n=1.007638-0.053592 \mathrm{i}$, and b), c), d) and e) the shifts for a set of magnetised surfaces with $Q=0.0215 e^{\mathrm{i}  0.073}$ and $n=1.007638-0.053592 \mathrm{i}$. Shifts are given in units of $k^{-1}$ on the $y$-axis and angle of incidence $\theta_0$ in radians on the $x$-axis.}
        %with a pseudo-Brewster's angle resonance at $\theta_0=\arctan{n}$.}%\ref{fig:figf} is to be contrasted with \ref{fig:figa}, where it can be seen both incident polarisations, for magnetisation in all directions produce shifts of the order $Q$ for all analyser polarisations. }
		\label{fig:vortdiff}
	\end{figure}

It immediately follows from the new functional form of $\mathbf{R}_\text{GH}$ and $\mathbf{R}_\text{IF}$ for homogeneous magnetic surfaces, that the magnetisation of the surface plays a role in changing the vortex centroid shift from that of the dielectric case. To analyse this role and refine the contributions from each direction of magnetisation, we plot the vortex centroid shift Eq.~(\ref{reflected_scalar_matrixform}) for different homogeneous magnetisation directions for direct comparison with the dielectric behaviour \cite{gotte_spin-orbit_2013} in Figure \ref{fig:vortdiff}. We chose the refractive index $n=1.007638-0.053592 \mathrm{i}$ from \cite{vimal_theoretical_2022, henke_x-ray_1993} and the magneto-optic constant $Q=0.0215 e^{\mathrm{i}  0.073}$ from \cite{piovera_c_ultrafast_2013, w_voigt_magneto-_1908} to correspond to the 3p-edge of Fe at $52.8$eV as is probed in the recent MHD experiments \cite{fanciulli_magnetic_2022}.

The vortex centroid shift, the magnitude of which is plotted in Figure \ref{fig:vortdiff} first for a homogeneously $p$ polarised beam incident on a dielectric surface, then on a selection of homogeneously magnetised surfaces, is given for a range of analyser polarisations $\mathbf{F}$ indicated by the coloured plots.

We first address the features of the dielectric case, also presented in \cite{gotte_spin-orbit_2013} for an ideal dielectric without optical absorption. For an ideal dielectric and an input $p$ polarised beam we would expect a shift resonance around Brewster's angle $\theta_B$, the angle of which can be related to the real refractive index $n$ by $\theta_B=\arctan(n)$. Brewster's angle can be understood as the angle at which Kerker's condition is fulfilled and the reflection coefficient $r_p$ goes to zero at this angle \cite{paniagua-dominguez_generalized_2016}. In the case considered here, we instead have an optically absorbing dielectric. The reflection coefficient $r_p$ does not go exactly to zero, rather reaches a minimum of reflection which can be found by solving $\partial \, \abs{r_p}^2 / \partial \, \theta_0=0$ \cite{akimoto_brewster_1992, humphreys-owen_comparison_1961}, where this minimum of reflection leads to a regularised shift resonance, a pseudo-Brewster's angle. As the dielectric surface we consider is not strongly absorbing, this pseudo-Brewster's peak falls roughly where the real $n$ Brewster's angle falls, around $\theta_0\approx0.79$ in Figure \ref{fig2:fige}. We also note for the dielectric case, that at normal incidence where $\theta_0=0$, there is no shift, and for the orthogonal analysed polarisation, $s$ shown in green, there is no shift throughout as to first order in $k$, the exactly orthogonal polarisation goes to zero intensity. 

We find that homogeneously magnetised surfaces exhibit a number of novel features. Starting with transverse magnetisation $m_t$ in Figure \ref{fig2:figd}, which was shown to modulate the $p$-channel amplitude in $R_\text{GH}$ and the spin-orbit term in $R_\text{IF}$ in Eq. (\ref{GH IF shifts magnetic}), we can see that the presence of the magnetisation shifts the pseudo-Brewster's angle to $\theta_0\approx0.74$, as well as enhancing the shifts observed. 

For longitudinal magnetisation $m_l$ in Figure \ref{fig2:figc}, which was shown to break the diagonal/anti-diagonal symmetry of both $\mathbf{R}_\text{GH}$ and $\mathbf{R}_\text{IF}$ through polarisation mixing, shows a splitting of the pseudo-Brewster's peak for different analyser polarisations, as well as a non-zero shift for the orthogonal polarisation $s$. The splitting into the three distinct peaks centred around the pseudo-Brewster's peak, blue ($45^\circ$), magenta ($p$), yellow ($135^\circ$), is unique for the polarisation mixing magnetisation terms $m_l$ and $m_p$. The polarisation mixing behaviour allows for a small portion of $p$ to be mixed to $s$ at all angles, on the order of the magneto-optic constant $Q$. When the intensity of the post-selected polarisation is very low with respect to the input intensity, this can produce very large vortex shifts, detailed with analogy to weak measurements in \cite{dennis_analogy_2012}. For the realistic value of $n\approx1$, the orthogonal shift is non-zero, but not significantly enhanced above other analysed polarisations.

Polar magnetisation $m_p$, also contributes to polarisation mixing, and also shows non-zero shifts for the orthogonal polarisation (Figure \ref{fig2:figb}) like longitudinal magnetisation. However, there is another effect distinct from dielectric behaviour (Figure \ref{fig2:fige}), in that there is a non-zero shift at $\theta_0=0$, or normal incidence, unique to polar magnetisation (Figure \ref{fig2:figb}). As the polar magnetisation at normal incidence causes a rotation in polarisation state and hence a change in polarisation \cite{freiser_survey_1968}, then at normal incidence to the surface this rotation of reflected polarisation is what facilitates the non-zero vortex centroid shift. This is not observed for longitudinal magnetisation at normal incidence, as longitudinal MOKE disappears for terms linear in $Q$ \cite{oppeneer_chapter_2001} when $\theta_0=0$, as evidenced by the functional form in Eq. \ref{all fresnel}, through $\sin\theta_0$ in the numerator of $r_{ps}^l$.

The result of non-zero magnetisation components in all directions is given in Figure \ref{fig2:figa}, which displays the components of all the magnetisations in an additive manner: the pseudo-Brewster's angle is shifted due to $m_t$, the resonance peak splits for blue ($45^\circ$), magenta ($p$), yellow ($135^\circ$) due to $m_p$ and $m_l$, there is the non-zero shift at normal incidence due to $m_p$, and there finally is the non-zero shift for orthogonally analysed polarisation $s$ in green. 

\section{Plane wave eigenpolarisations and modified non-Hermitian behaviour}
\label{sec:eigenpolarizations}

Eigenpolarisations along with their corresponding eigenvalues allow for the reflection properties of the surface to be characterised. If we consider the reflection matrix to be acting as a polarising element, we can define the polarising element as homogeneous when its eigenpolarisations are orthogonal, and inhomogeneous when non-orthogonal \cite{lu_homogeneous_1994}. 
%%%and in particular, can utilised to calculate the total phase acquired upon reflection. The total phase change, due to both dynamical phase from the path travelled and the Pancharatnam-Berry, or geometric phase acquired from reflection of the surface \cite{bliokh_geometric_2019, berry_adiabatic_1987}, can be calculated for any input polarisation with a complete treatment derived by Gutiérrez-Vega in \cite{gutierrez-vega_pancharatnamberry_2011}. The geometric phase is connected to changes in polarisation state, where the eigenpolarisations and eigenvalues are utilised to calculate this phase, and conversely, geometric phase polarimetry can be used to recover the eigenpolarisations and eigenvalues of the surface \cite{garza-soto_geometric-phase_2020}.

It is helpful at this point to introduce the idea of Hermitian and non-Hermitian matrices. Non-hermitian matrices have found application in the potential for enhanced sensing and for optics and photonics in polarisation splitting \cite{zyablovsky_pt-symmetry_2014, longhi_parity-time_2018, jiang_harnessing_2026, liu_metrology_2016}. Hermitian matrices posses a real eigenvalue spectrum and orthogonal eigenvectors. An optical counterpart can be constructed by considering plane wave reflection off of a perfect dielectric where $n>1$ and is real, such that the eigenvalues are always real and the eigenvectors are orthogonal. Non-hermitian matrices are the set of matrices which have a complex eigenvalue spectrum, and a corresponding set of non-orthogonal eigenvectors, which can be used to describe open quantum systems, where there is some non-unitary gain or loss with an environment \cite{el-ganainy_non-hermitian_2018}. 
In particular, the signature of non-Hermitian behaviour is the existence of exceptional points (EPs), which is the set of point(s) in parameter space where these eigenpolarisations become parallel at the same point the eigenvalues become degenerate, referred to as a non-Hermitian degeneracy \cite{miri_exceptional_2019, el-ganainy_non-hermitian_2018}. In this case, the polarising element, the reflection matrix, can be said to become degenerate \cite{savenkov_eigenanalysis_2007}.

%A specific class of non-Hermitian matrices, which posses a transition from real to complex eigenvalues across the EP due to exactly balanced loss and gain, have been utilised to study PT symmetry, and were found to have applications in quantum mechanics, where the existence of real eigenvalues can be utilised in simulating open quantum systems with gain and loss between the two eigenstates \cite{croke_mathcalpt-symmetric_2015,lu_quantum_2025, zhao_magnetically_2025}.

The eigenpolarisations can be calculated as outlined around Eq.~\eqref{eigpol_seperate}, where the eigendecomposition of $\mathbf{R}_0$ gives the eigenpolarisations and values for an incident plane wave. For cases where we have non-Hermitian behaviour, the eigenpolarisations quoted are the right eigenpolarisations.
For dielectric surfaces, these are as expected, 

\begin{equation}\label{eigen dielectric}
    \begin{aligned}
  \lambda^{(e)}_{1} & =
r_p &\quad \lambda^{(e)}_{2} & =r_s,
\\
\\
\nu^{(e)}_{1} &=
 \begin{pmatrix}
  1\\0
\end{pmatrix}
 &\qquad
\nu^{(e)}_{2} &=\begin{pmatrix}
  0\\1  
\end{pmatrix},
\end{aligned}
\end{equation}
where we recover the $s$ and $p$ polarisations as orthogonal eigenvectors and the eigenvalues as the corresponding Fresnel coefficients. This form is simple by construction, as the reflection matrix is given in the $sp$ basis. For lossy dielectrics with complex $n$, the eigenvalues on the diagonal are always complex, hence the plane wave reflection matrix non-Hermitian, without the possibility of an EP, as the eigenvectors can never coalesce. 
 
The second result of this work is the plane wave eigendecomposition for a homogeneous magnetic surface:  

\begin{equation}\label{eigen magnetic}
\begin{aligned}    
  \lambda^{(m)}_{1} &=
\frac{1}{2}\left(r_p +r_s + m_t r_0^t r_p-\sqrt{\chi}\right) &\quad \lambda^{(m)}_{2}&=\frac{1}{2}\left(r_p +r_s + m_t r_0^t r_p+\sqrt{\chi}\right) ,
\\
\\
\nu^{(m)}_{1} &=
    \begin{pmatrix}
  -\left(\frac{r_p -r_s + m_t r_0^t r_p-\sqrt{\chi}}{2(m_l r_{ps}^l-m_p r_{ps}^p)}\right)\\1  
\end{pmatrix}
 &\quad
\nu^{(m)}_{2} &=
    \begin{pmatrix}
  -\left(\frac{r_p -r_s + m_t r_0^t r_p+\sqrt{\chi}}{2(m_l r_{ps}^l-m_p r_{ps}^p)}\right)\\1  
\end{pmatrix},
\\
\end{aligned}
\end{equation}
where $\chi=-4m_l^2r_{ps}^{l 2}+4m_p^2r_{ps}^{p2} +(r_p-r_s+m_t r_0^t r_p )^2 $. Note that for some surface magnetisations, the eigenpolarisations are no longer orthogonal for plane waves, and hence can be utilised to explore non-Hermitian behaviour, particularly the existence of an EP for the plane wave reflection case.  

%In the case of homogeneous magnetic surfaces, for realistic values of $Q\ll1$ and complex $n$, we do not observe the PT symmetry transition, which requires an exact balance of gain and loss between the two eigenpolarisations to demonstrate the transition from all real to complex eigenvalues at the EP \cite{ozdemir_paritytime_2019}. However, we do not think it would be impossible to engineer a surface with the current homogeneous magnetic reflection matrix for which the plane wave EP displays this PT symmetry transition. 
%for example for a metasurface with real $n$, for $m_t=0$ where $(-m_l r_{ps}^l+m_p r_{ps}^p)=(m_l r_{ps}^l+m_p r_{ps}^p)^*$, $^*$ denotes the complex conjugate such that the gain and loss where balanced are the diagonals are real. 

%Despite the absence of the PT symmetry transition, 
We propose that the existence of the EP for plane wave reflection itself offers a way to probe non-Hermitian physics. It has been shown for scattering matrices for the GH shift from lossy thin-films that around an EP in the GH shift, the shift offers enhanced sensing capability \cite{zhao_giant_2018}, our method of using reflection matrices for magnetic surfaces would be a novel application with similar motivating principles in this case in that the rapid change of eigenpolarisation around the EP would be the generator of the sensitivity. 

We postulate further that singularimetry could be utilised to probe around the plane wave exceptional point and hence non-Hermitian systems, as the beam has a narrow spread around the EP, the rapid change in plane wave eigenpolarisations can be accessed directly making it easier to access the EP behaviour with the wider range of incidence angle provided by the beam, with potential applications as a magnetometer both from the vortex positions after reflection and the modified polarisation structure of the reflected beam which can be numerically calculated as shown in Section \ref{sec:vortex_eigenpolarizations}. 

For dielectric reflection, for both real and complex $n$, the plane wave eigenvectors are always orthogonal, so the dielectric $\mathbf{R}_0$ is said to be homogeneous \cite{gutierrez-vega_how_2020,gutierrez-vega_pancharatnamberry_2011}.
%For all other cases where the eigenpolarisations are non-orthogonal, the Jones matrix is then said to be inhomogeneous or degenerate \cite{lu_homogeneous_1994}. 
%Characterising this degree of inhomogeneity, a measure of how parallel the eigenpolarisations are, is required to characterise the reflection properties of the surface, as total phase changes are generated by both the eigenvalues and the inhomogeneity of the Jones matrix \cite{lu_homogeneous_1994, gutierrez-vega_how_2020}. 
For a homogeneously magnetised surface, the form of Eq. (\ref{eigen magnetic}) is valid for magnetisations where both $m_l$ and $m_p$ are not both equal to zero. In the case of only transverse magnetisation $m_t$, that is $m_l=m_p=0$, we recover the $sp$ polarisation basis of the dielectric case, Eq. ~\eqref{eigen dielectric}, with the modification to the $p$ eigenvalue, $\lambda_{(e)1}=r_p +m_t r_0^tr_p$. For solely transverse magnetisation, the Jones matrix is therefore still homogeneous, and the non-orthogonality for the plane wave eigenpolarisations is generated by the off-diagonal magnetisation components $m_p$ and $m_l$. 

Non-Hermitian behaviour has been explored for the dielectric eigenpolarisations of the shift matrices  $\mathbf{R}_\text{IF}$, and has opened the avenue of exploring non-Hermitian behaviour with the IF shift for reflection of beams from dielectric surfaces with real $n$ \cite{el-ganainy_non-hermitian_2018, gotte_eigenpolarizations_2014, modak_inhomogeneous_2024}. For dielectric surfaces, it can be seen from the form of the plane wave eigenpolarisations that there is no possible combination of angle or refractive index where both the eigenvalues can become degenerate, at the same time the eigenpolarisations are parallel. This is because the eigenpolarisations are not dependent $\theta_0$ or $n$ and the degeneracy at $\theta_0=0$ for $r_p$ and $r_s$ is a normal degeneracy, not an EP. Hence for dielectric surfaces, the only possible way to access possible EPs is to move to the first order in $\delta$ and look at the eigenpolarisations of the vortex beam and its related IF shifts, detailed in the following, Section \ref{sec:vortex_eigenpolarizations}. 

\section{Unit order scalar vortex beam eigenpolarisations for dielectric and homogeneous magnetic surfaces}
\label{sec:vortex_eigenpolarizations}

As with singularimetry, the reflection of vortex beams requires a more sophisticated treatment. The vortex beam is comprised of a bundle of plane waves, which individually all have separate individual $sp$ eigenbasis which are not identical for all plane waves comprising the beam, and hence are not the eigenbasis of the vortex beam itself. This is a direct consequence of both the finite physical extent of the beam, where each plane wave is incident at a local angle of incidence as in Eq.~\eqref{local theta}, and the presence of the topological object, the unit order phase vortex. The vortex introduces additional topological complexity in that the phase, and hence resulting polarisation must re-arrange around the vortex when reflected, affecting which eigenpolarisations are selected for the vortex beam.  

\begin{figure}[t] % t for position at the top of the current page; b for position at the bottom; p for new page
		\centering
        \begin{subfigure}[b]{0.34\linewidth} % Fig (a)
			\includegraphics[width=\linewidth]{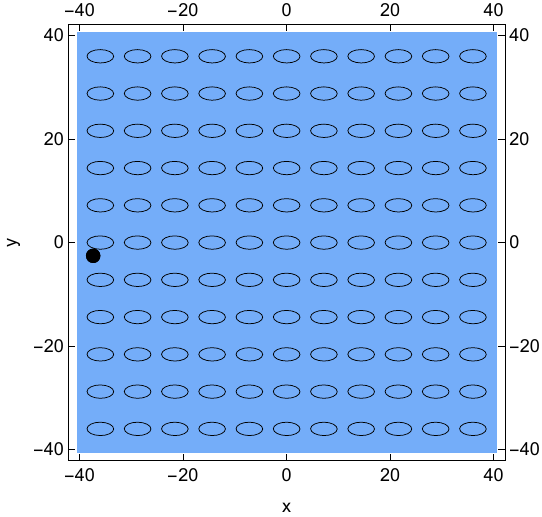}
			\caption{$\nu^{(e)}_1$ with vortex centroid in black.}
			\label{fig3:figa}
		\end{subfigure}
			\hspace{20pt}   % Space between the figures
		\begin{subfigure}[b]{0.34\linewidth} % Fig (b)
			\includegraphics[width=\linewidth]{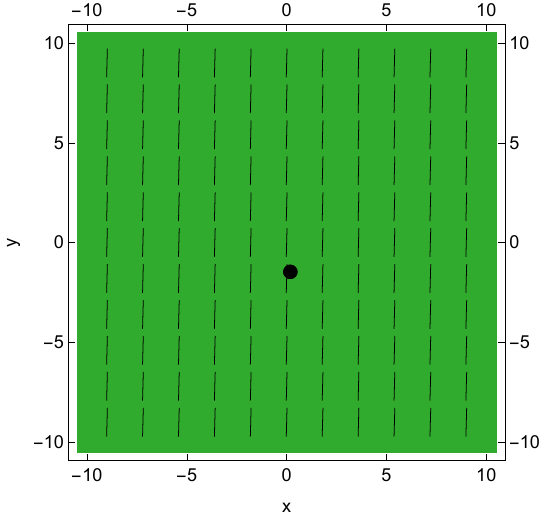}
			\caption{$\nu^{(e)}_2$ with vortex centroid in black.}
			\label{fig3:figb}
		\end{subfigure}
		  \begin{subfigure}[b]{0.34\linewidth} % Fig (e)
			\includegraphics[width=\linewidth]{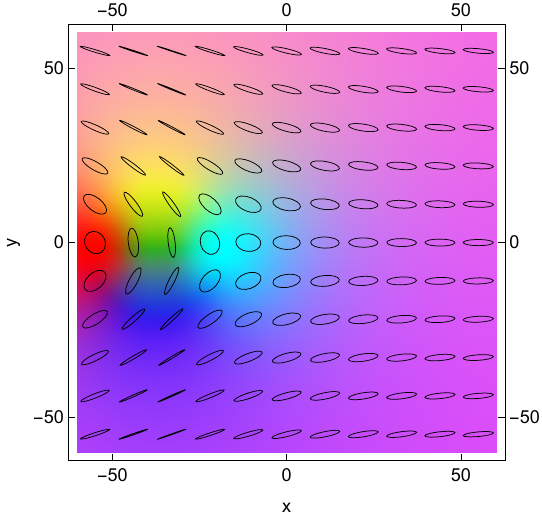}
			\caption{$p$ polarised beam on a dielectric surface.}
			\label{fig3:figc}
		\end{subfigure}
        	\hspace{20pt}   % Space between the figures
		\begin{subfigure}[b]{0.34\linewidth} % Fig (b)
			\includegraphics[width=\linewidth]{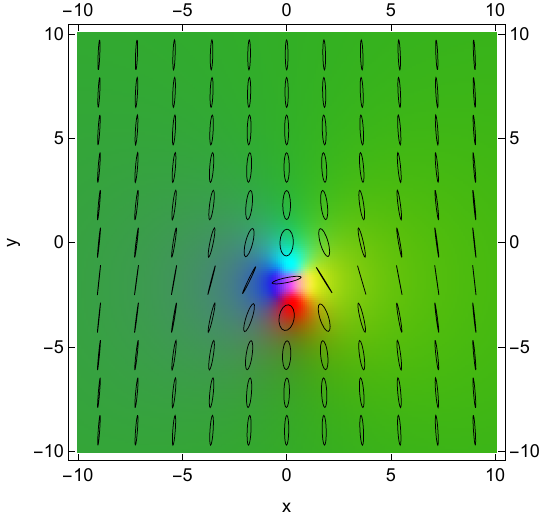}
			\caption{$s$ polarised beam on a dielectric surface.}
			\label{fig3:figd}
		\end{subfigure}
        \begin{subfigure}[b]{0.34\linewidth} % Fig (b)
			\includegraphics[width=\linewidth,]{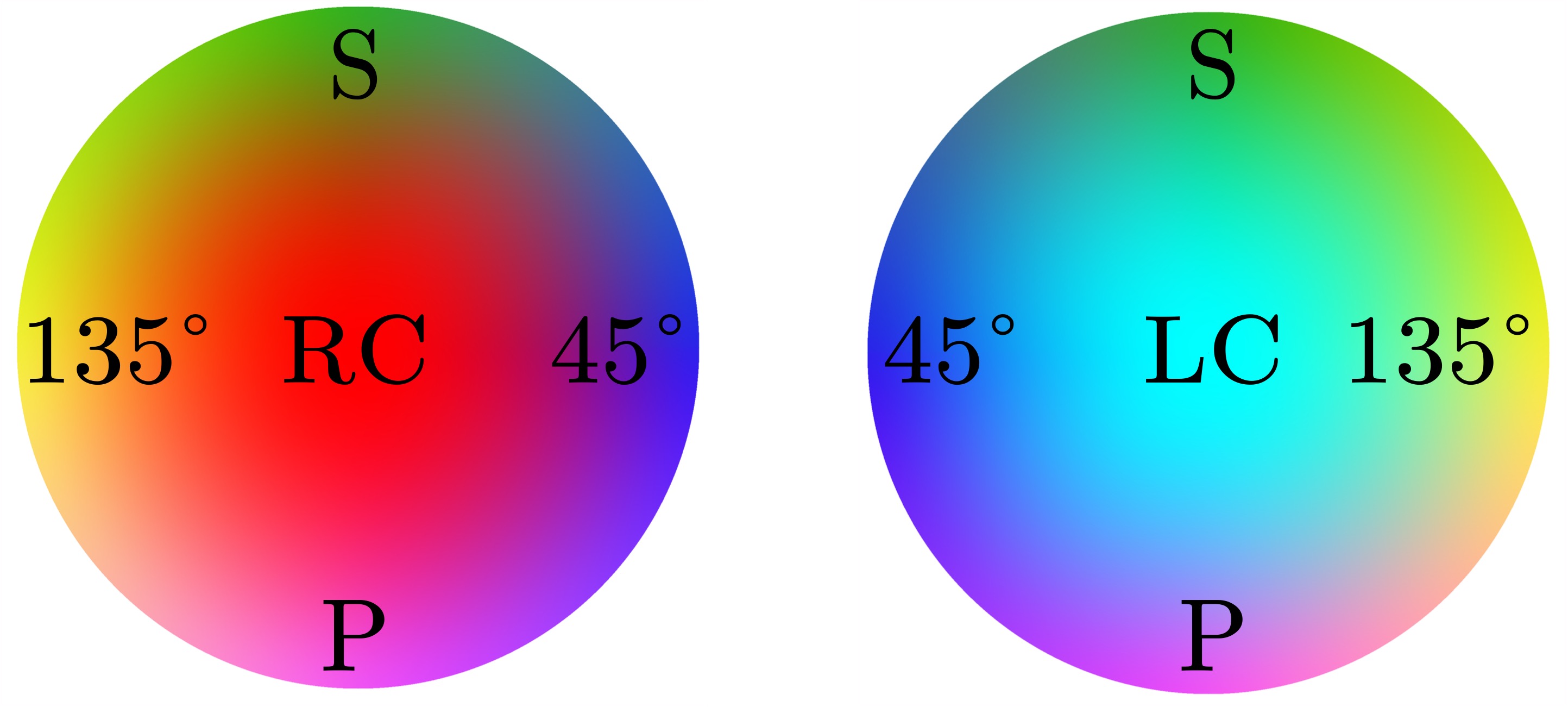}
			\caption{North and south projections of the Poincaré sphere for polarisation shading. }
			\label{fig3:fige}
		\end{subfigure}
		\caption{Plots of eigenpolarisations for a dielectric surface, as polarisation ellipses in black, and colour coded polarisation shading with mapping to the Poincaré sphere shown in the reference image \ref{fig3:fige}. Eigenpolarisations display homogeneous spatial structure after reflection in \ref{fig3:figa} and \ref{fig3:figb}. $p$ polarised beam showing spatially varying reflected structure in \ref{fig3:figc}, with vortex centroid now analyser dependent highlighted by the rainbow of polarisations around the vortex. Similarly for \ref{fig3:figd} with $s$ polarisation.  All axes in units of $k^{-1}$, with $n=1.007638-0.053592 \mathrm{i}$ and $\theta_0$ was chosen to fall close to the pseudo-Brewster's angle at $\theta_0\approx45^\circ\approx0.79$.}
		\label{fig:beampoldia}
	\end{figure}

The third result of this work, the eigenpolarisations of a unit order vortex beam are calculated as discussed around Eq. \eqref{eigpol_original}, with the set of two transverse eigenpolarisations for a flat dielectric surface given in Figure \ref{fig:beampoldia}, along with both $s$ and $p$ polarised vortex beams to show how the polarisation rearranges around the vortex when a non-eigenpolarisation is incident on the surface.

From Figure \ref{fig3:figa}, it can seen that the first eigenpolarisation for a unit order vortex beam for a dielectric surface is the perturbed $p$ eigenpolarisation, $\nu^{(e)}_1$, which for a vortex beam is significantly more circularly polarised. The second eigenpolarisation for a vortex beam, Figure \ref{fig3:figb}, is the perturbed $s$ eigenpolarisation, $\nu^{(e)}_2$, which we find tends to be less perturbed from the plane wave eigenpolatisation than the first eigenpolarisation. 

Figure \ref{fig:beampoldia}c and d show the expected behaviour of any other input polarisation which does not belong to the set of eigenpolarisations of the vortex beam. Upon reflection, the polarisation re-arranges around the phase vortex, where it can be seen that all polarisation states on the Poincaré sphere are populated as a result of the topological character of the vortex, through the presence of all colours on the Poincaré sphere (the polarisation ellipses are too coarse grained to see from these alone). In general, the position of the vortex centroid, Eq. \eqref{reflected_scalar_matrixform}, is dependent both analyser polarisation $\mathbf{F}$ and input polarisation $\mathbf{E}$, as shown in Figure \ref{fig:vortdiff} where the input homogeneous $p$ polarisation is not an eigenpolarisation of the vortex beam. As the reflected polarisation is no longer spatially homogeneous, different analyser $\mathbf{F}$ results in a shifted intensity centroid, as the distribution of intensity is no longer identical across all polarisations. For the special case of eigenpolarisations of the beam however, the vortex centroid, and hence intensity centroid, is independent of the analyser $\mathbf{F}$, as the eigenpolarisation is spatially homogeneous.

\begin{figure}[t] % t for position at the top of the current page; b for position at the bottom; p for new page
		\centering
		  \begin{subfigure}[b]{0.21\linewidth} % Fig (c)
            \includegraphics[width=\linewidth,clip]{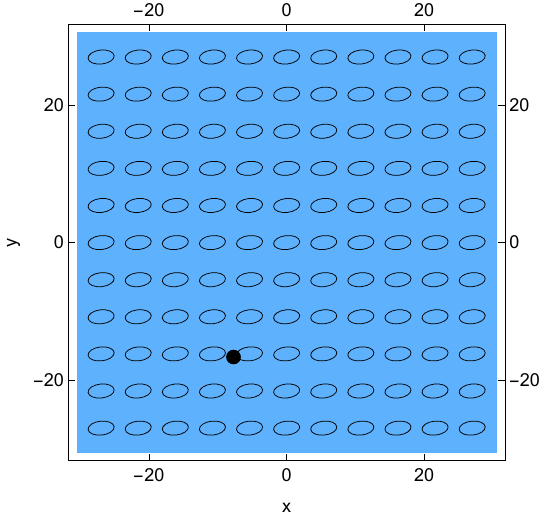}
			\caption{$\nu^{(m)}_{1}$, where $m_{p,l,t}=1/ \sqrt{3}$.}
			\label{fig4:figa}
		\end{subfigure}
			\hspace{10pt}   % Space between the figures
		\begin{subfigure}[b]{0.21\linewidth} % Fig (d)
			\includegraphics[width=\linewidth,clip]{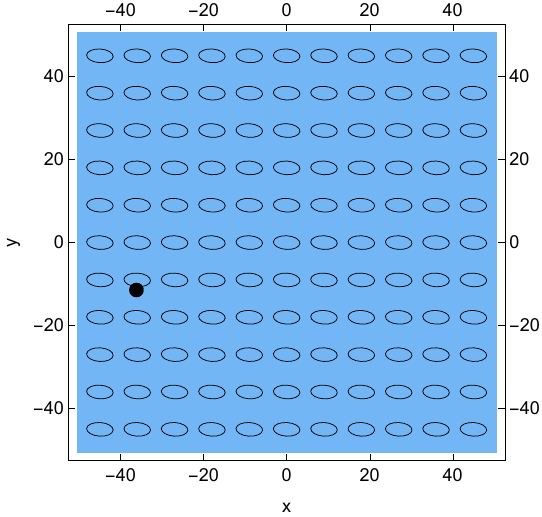}
			\caption{$\nu^{(m)}_{1}$, where \,$m_p=1$.}
			\label{fig4:figb}
		\end{subfigure}
        \hspace{10pt}
		\begin{subfigure}[b]{0.21\linewidth} % Fig (f)
			\includegraphics[width=\linewidth,clip]{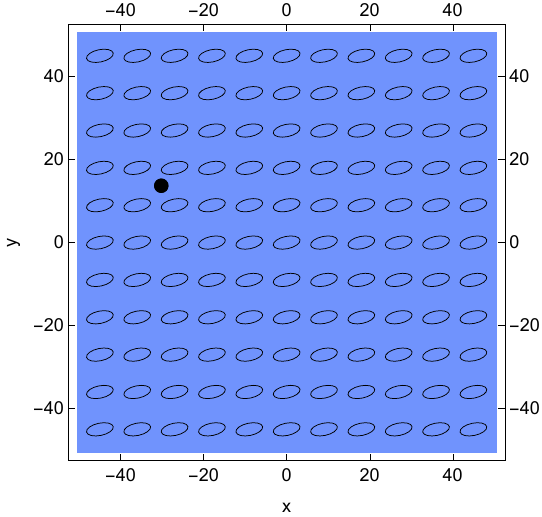}
			\caption{$\nu^{(m)}_{1}$, where \, $m_l=1$. }
			\label{fig4:figc}
            \end{subfigure}
            \hspace{10pt}   % Space between the figures
		\begin{subfigure}[b]{0.21\linewidth} % Fig (d)
			\includegraphics[width=\linewidth,clip]{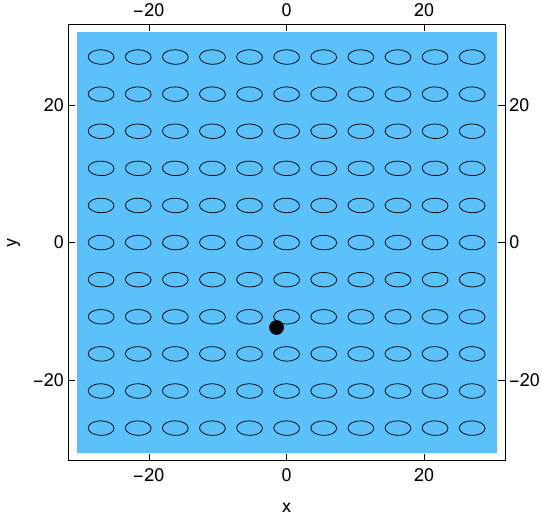}
			\caption{$\nu^{(m)}_{1}$, where \, $m_t=1$.}
			\label{fig4:figd}
		\end{subfigure}
         \begin{subfigure}[b]{0.21\linewidth} % Fig (c)
            \includegraphics[width=\linewidth,clip]{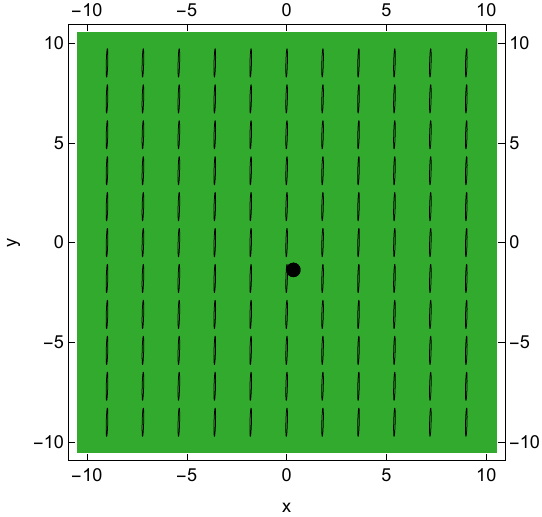}
			\caption{$\nu^{(m)}_{2}$, where \, $m_{p,l,t}=1/ \sqrt{3}$.}
			\label{fig4:fige}
		\end{subfigure}
			\hspace{10pt}   % Space between the figures
		\begin{subfigure}[b]{0.21\linewidth} % Fig (d)
			\includegraphics[width=\linewidth,clip]{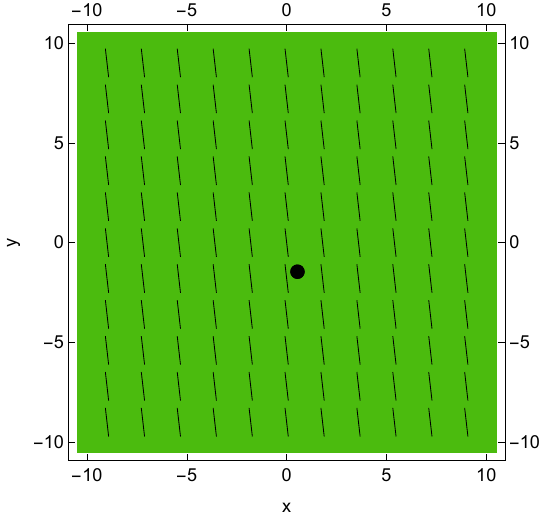}
			\caption{$\nu^{(m)}_{2}$, where \, $m_p=1$.}
			\label{fig4:figf}
		\end{subfigure}
        \hspace{10pt}
		\begin{subfigure}[b]{0.21\linewidth} % Fig (f)
			\includegraphics[width=\linewidth,clip]{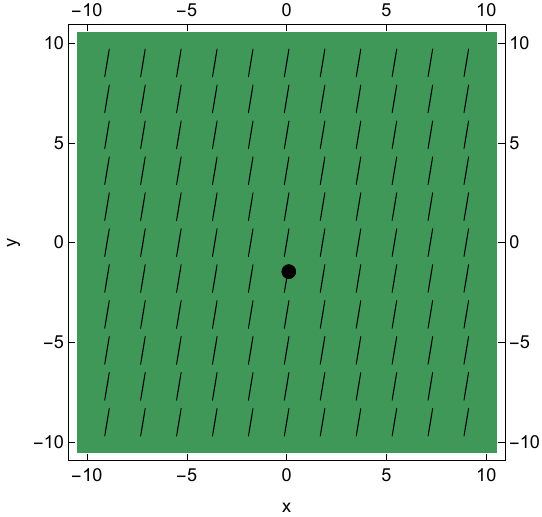}
			\caption{$\nu^{(m)}_{2}$, where \, $m_l=1$. }
			\label{fig4:figg}
            \end{subfigure}
            \hspace{10pt}   % Space between the figures
		\begin{subfigure}[b]{0.21\linewidth} % Fig (d)
			\includegraphics[width=\linewidth,clip]{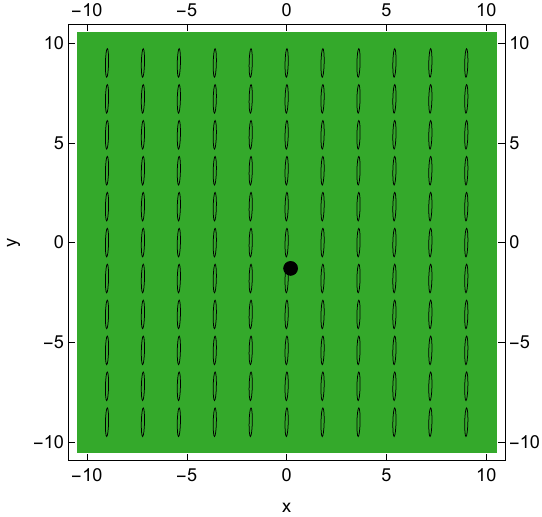}
			\caption{$\nu^{(m)}_{2}$, where \, $m_t=1$.}
			\label{fig4:figh}
        \end{subfigure}
		\caption{Plots of eigenpolarisations for a homogeneous magnetic surface, with polarisation displayed as in Figure \ref{fig:beampoldia}. All axes in units of $k^{-1}$, with $\theta_0=0.79$, refractive index $n=1.007638-0.053592 \mathrm{i}$. Note larger axes area for both $m_p$ and $m_l$ eigenpolarisations indicating the larger shifts for these magnetisations when $m_t$ is absent.}
		\label{fig:beampolmag}
	\end{figure}

It is illustrated in Figure \ref{fig:beampoldia} that homogeneous eigenpolarisations of $l=\pm1$ vortex beams exist for a dielectric surface. Homogeneous transverse eigenpolarisations also exist for  homogeneous magnetic surfaces, illustrated in Figure \ref{fig:beampolmag}, solved for in the same manner as for the dielectric case using the magnetic counterpart of Eq. \eqref{eigpol_seperate}. The specific surface parameters were chosen to be close to pseudo-Brewster's angle shown in Figure \ref{fig2:figa}. The eigenpolarisations themselves are dependant on $\theta_0$ and $n$, and give different solutions for changed values. The eigenpolarisations are special in that they select the $\mathbf{E}_i$ which gives us a  homogeneously polarised reflected beam, for all other inputs tested the reflected beam becomes inhomogeneously polarised, regardless of if the input polarisation $\mathbf{E}_i$ is homogeneous itself. 
%We again consider the set of input polarisations $\mathbf{E}_i$ rather than the experimental input polarisations $\mathbf{E}$ as it allows us to 

The non-Hermitian character, and appearance of exceptional points, for the IF shift matrix $\mathbf{R}_\text{IF}$ in particular are well studied for dielectric surfaces with real $n$ and the point at Brewster's angle is associated with the PT symmetry phase change \cite{gotte_eigenpolarizations_2014, modak_inhomogeneous_2024}. The resulting eigenvalues determine the largest possible IF shift for the corresponding eigenpolarisation \cite{gotte_eigenpolarizations_2014}. For the GH shift, $\mathbf{R}_\text{GH}$, for dielectric surfaces the eigenpolarisations are just the $sp$ basis as can be seen from the form in equation Eq. \eqref{GH IF shifts dielectric}, and hence in the reflection matrix formalism cannot support an EP. For lossless dielectric surfaces, Brewster's angle is an exceptional point for the IF vortex shift, where the eigenpolarisations are parallel with degenerate eigenvalues, transitioning from imaginary eigenvalues for angular shifts below the resonance, and spatial shifts for real eigenvalues above. For our homogeneous magnetic case, using $Q\ll1$ and complex $n$, we find as expected that in general the eigenvalues for $\mathbf{R}_\text{IF}$ are complex both above and below pseudo-Brewster's angle, indicating both angular and spatial shifts.

%As with the plane wave eigenpolarisations, we do not observe the real to complex eigenvalue transition associated with PT symmetry for the EP at pseudo-Brewster's angle, as we again do not have the balanced gain and loss for the two eigenpolarisations. 
Given we also have off-diagonal terms in the $\mathbf{R}_\text{GH}$ matrix for homogeneous magnetic surfaces, Eq. \eqref{GH IF shifts magnetic}, the $sp$ polarisations are no longer the eigenpolarisations and there is the potential for EP here where it is previously not possible for the GH shift for dielectric surfaces with real $n$. For lossy dielectrics in metamaterials with complex $n$, a configurable EP allows for applications in sensing, where rapid change in phase leads to a change in sign of the GH shift \cite{zhao_giant_2018}. 
%To fully probe the eigenvalues and eigenpolarisations directly on the EP at Brewster's angle for lossless surfaces with real $n$, we would need to continue our expansion to second order $\delta$ as in \cite{gotte_limits_2013} for dielectric surfaces as alluded to in the analysis of the Brewster EP for the IF shift in \cite{modak_inhomogeneous_2024} to normalise the Brewster's shift peaks. Such treatment is not required for lossy materials with complex $n$, as the defective point within the now pseudo-Brewster's peak, normally where $r_p$ goes exactly to zero, is replaced with a non-zero value, so that the reflection matrix can be exactly evaluated at these EPs. 

As we are interested here in the eigenpolarisations of the entire reflection matrix, we calculate the eigenpolarisations as detailed in Eq. \eqref{eigpol_seperate}. We find in general these eigenpolarisations are similar to the comparable ones in the dielectric case, in that the magnetisation acts almost in a perturbation like manner to change the eigenpolarisations close to the pseudo-Brewster's angle. We note for $\nu^{(m)}_1$ eigenpolarisations where there is $m_t$ magnetisation,  Figure \ref{fig4:figa} and Figure \ref{fig4:figd}, the eigenpolarisation is most similar to the dielectric case Figure \ref{fig3:figa}, with a similar vortex centroid shift. For $m_l$ and $m_p$, Figures \ref{fig4:figb} and \ref{fig4:figc}, we see significantly enhanced vortex centroid shifts as expected from the larger peaks at this angle for comparable polarisations for  Figure \ref{fig2:figb} and  Figure \ref{fig2:figc}. A less dramatic effect is present for $\nu^{(m)}_2$ eigenpolarisations, where the $m_t$ present cases, Figures \ref{fig4:fige} and \ref{fig4:figh}, the $\nu^{(m)}_2$ eigenpolarisations display greater ellipticity in comparison to the dielectric case in Figure \ref{fig3:figb}. 

\begin{figure}[t] % t for position at the top of the current page; b for position at the bottom; p for new page
		\centering
		  \begin{subfigure}[b]{0.21\linewidth} % Fig (c)
            \includegraphics[width=\linewidth,clip]{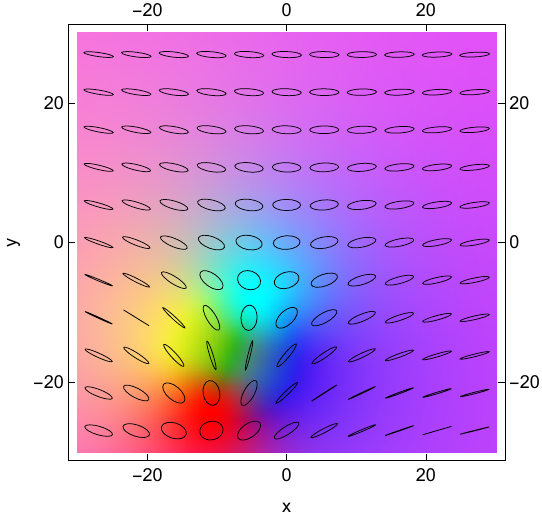}
			\caption{$p$-polarised, for $m_{p,l,t}=1/ \sqrt{3}$.}
			\label{fig5:figa}
		\end{subfigure}
			\hspace{10pt}   % Space between the figures
		\begin{subfigure}[b]{0.21\linewidth} % Fig (d)
			\includegraphics[width=\linewidth,clip]{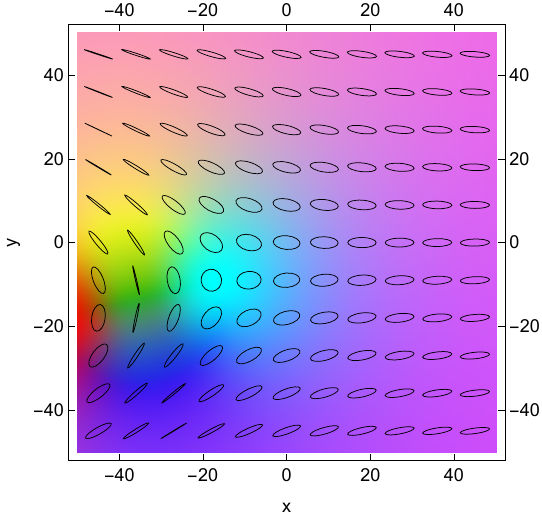}
			\caption{$p$-polarised, for $m_p=1$.}
			\label{fig5:figb}
		\end{subfigure}
        \hspace{10pt}
		\begin{subfigure}[b]{0.21\linewidth} % Fig (f)
			\includegraphics[width=\linewidth,clip]{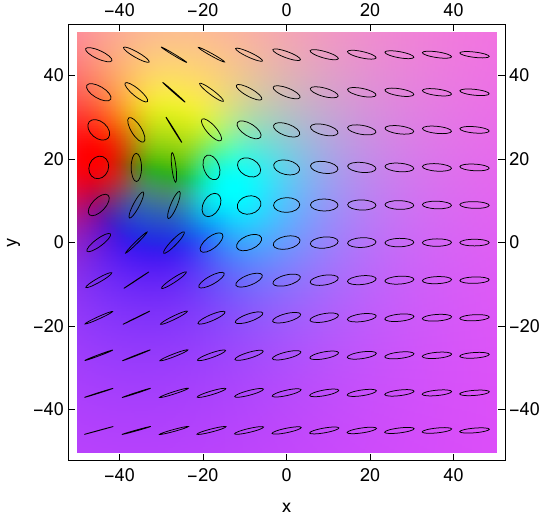}
			\caption{$p$-polarised, for $m_l=1$. }
			\label{fig5:figc}
            \end{subfigure}
            \hspace{10pt}   % Space between the figures
		\begin{subfigure}[b]{0.21\linewidth} % Fig (d)
			\includegraphics[width=\linewidth,clip]{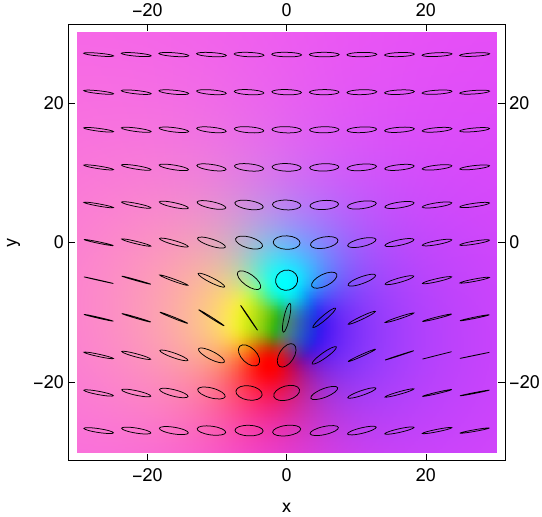}
			\caption{$p$-polarised, for $m_t=1$.}
			\label{fig5:figd}
		\end{subfigure}
         \begin{subfigure}[b]{0.21\linewidth} % Fig (c)
            \includegraphics[width=\linewidth,clip]{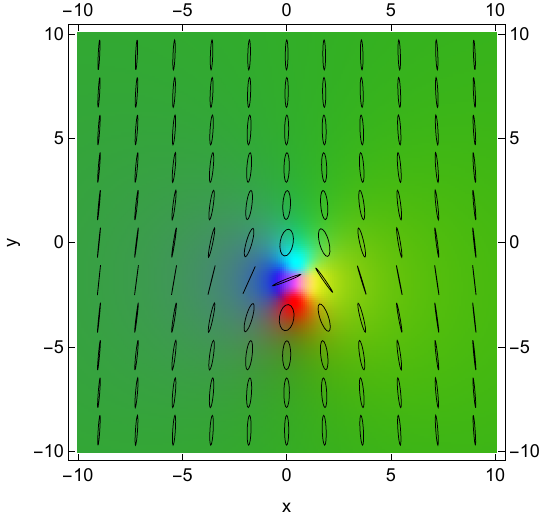}
			\caption{$s$-polarised, for $m_{p,l,t}=1/ \sqrt{3}$.}
			\label{fig5:fige}
		\end{subfigure}
			\hspace{10pt}   % Space between the figures
		\begin{subfigure}[b]{0.21\linewidth} % Fig (d)
			\includegraphics[width=\linewidth,clip]{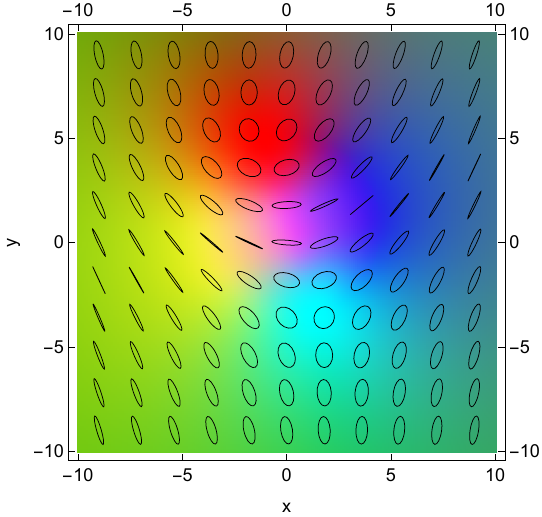}
			\caption{$s$-polarised, for $m_p=1$.}
			\label{fig5:figf}
		\end{subfigure}
        \hspace{10pt}
		\begin{subfigure}[b]{0.21\linewidth} % Fig (f)
			\includegraphics[width=\linewidth,clip]{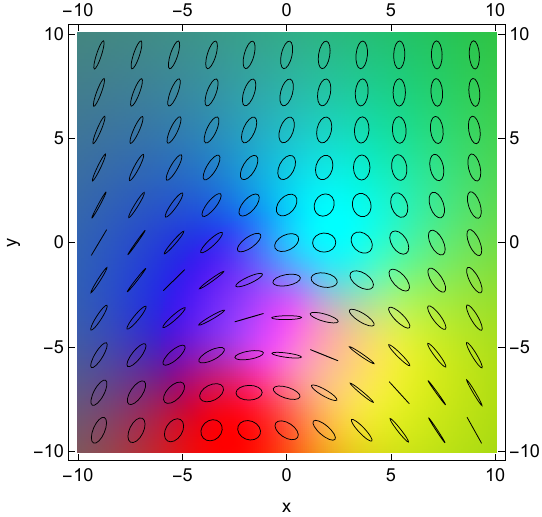}
			\caption{$s$-polarised, for $m_l=1$. }
			\label{fig5:figg}
            \end{subfigure}
            \hspace{10pt}   % Space between the figures
		\begin{subfigure}[b]{0.21\linewidth} % Fig (d)
			\includegraphics[width=\linewidth,clip]{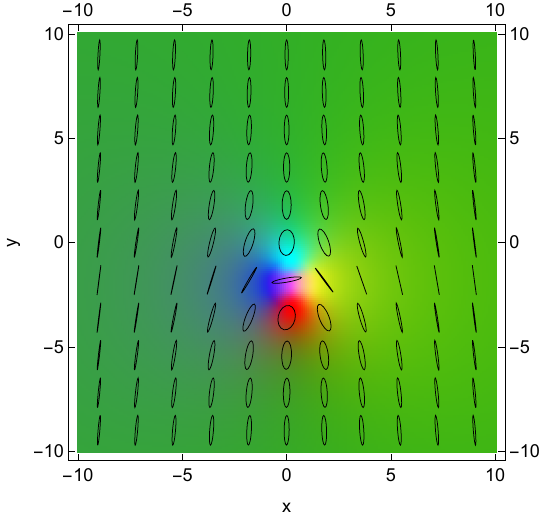}
			\caption{$s$-polarised, for $m_t=1$.}
			\label{fig5:figh}
		\end{subfigure}
        
		\caption{Plots of $p$ and $s$-polarised beams incident on a spatially homogeneous magnetic surface, with polarisation displayed as in Figure \ref{fig:beampoldia}. All axes in units of $k^{-1}$, with $\theta_0=0.79$, refractive index $n=1.007638-0.053592 \mathrm{i}$. Note larger axes area for both $m_p$ and $m_l$ eigenpolarisations in Fig \ref{fig5:figa} to Fig \ref{fig5:figd} indicating the larger shifts for these magnetisations when $m_t$ is absent.}
		\label{fig:magsp}
	\end{figure}

The non-eigenpolarisation inputs shown in Figure \ref{fig:magsp} are helpful in demonstrating further the differences between the magnetic and dielectric case, as well as the differences between magnetisation components. For $s$ polarised beams, the $m_l$ and $m_p$ cases in Figures \ref{fig5:figf} and \ref{fig5:figg} display significantly increased spot size, while the $m_t$ component, Figure \ref{fig5:figh} only slightly increases spot size while the other magnetisation components rotate the spot (note for example the right circularly polarised component is now rotated clockwise).

%The $s$-like eigenpolarisation, Figure \ref{fig4:figb}, is more similar to the dielectric case, where again for a non-eigenpolarisation input, Figure \ref{fig4:figc}, we see both the enhanced shift, but with a similar rearranging of polarisation around the vortex as with the dielectric cases, albert with a rotating of the ``ring'' of polarisations (note for example the right circularly polarised component is now rotated clockwise). For combinations of angle $\theta_0$ and refractive index $n$ where the vortex centroid shifts are similar for the magnetic and dielectric cases, the eigenpolarisations can also be quite similar, with differences in vortex centroid shift on the order of the magneto-optical constant $Q$. 
For non-homogeneous magnetic surfaces, it is expected that the eigenpolarisations for a unit order vortex beam will not be spatially homogeneous, rather should reflect the symmetries of the non-homogeneous surface magnetisation. Similarly, for an $l=\pm2$ order vortex beam, eigenpolarisations are expected to be non-homogeneous for both dielectric and homogeneous magnetic surfaces as the splitting of the higher order vortex into unit vortices upon reflection, would allow for non-homogeneous polarisation structures between the unit order vortices. A notable exception would be if the second order and higher expansions of the reflection matrix $R$ commute with Eq. \eqref{eigpol_original}, in which case they should share eigenpolarisations and hence $l=\pm2$ could from a mathematical standpoint possess spatially homogeneous eigenpolarisations and similarly higher order vortex beams by the same argument.

%shows how for a homogeneous magnetic surface, the reflected vortex beam can have a set of spatially homogeneous transverse eigenpolarisations, in a similar manner to a vortex beam reflected from a dielectric surface, and at some combinations of angle and refractive index these eigenpolarisations can be quite similar, on the order of $Q$ and for other combinations these differences are more pronounced. In all cases, they allow for a complete description of the resulting total phase and amplitude change after reflection from the homogeneously magnetised surface. 
\section{Discussion and Conclusions}
\label{sec:discussion}

In summary, we have utilised the singularimetry approach to analyse homogeneous magnetic surfaces, where we have found the analytic forms of the Artmann operators for a homogeneously magnetised surface. These allow for the calculation of the GH and IF shifts for unit order vortex beams. The shifts for homogeneous surfaces reveal additional reflection behaviour for magnetised surfaces. We introduce the possibility of exceptional points for  $\mathbf{R_\text{GH}}$ for homogeneous magnetic surfaces and for plane waves from homogeneous magnetic surfaces \cite{modak_inhomogeneous_2024, el-ganainy_non-hermitian_2018, gotte_eigenpolarizations_2014}.

The polarisation mixing terms, associated with longitudinal $m_l$ and polar $m_p$ magnetisation, introduce new behaviour when considering an orthogonal analyser polarisation $\mathbf{F}$. The mixing of a small fraction of incident beam intensity to the orthogonal polarisation allows for both non-zero shifts.
%, and for $n\gg1$ significantly enhanced orthogonal shifts, analogous to the weak measurement effect \cite{dennis_analogy_2012}. 
The extension of these beam shift Artmann operators to homogeneous magnetic surfaces allows us to understand how these diffractive corrections to the laws of geometric optics are modified in the presence of polarisation mixing terms and magnetisation. Namely, the polarisation mixing terms breaking the diagonal/anti-diagonal symmetry observed for $\mathbf{R}_\text{GH}$ and $\mathbf{R}_\text{IF}$ respectively, and transverse magnetisation $m_t$ modulating the spin-orbit coupling. 

We found analytic expressions for the plane wave eigenpolarisations of a homogeneous magnetic surface in the $sp$-basis. By calculating the plane wave eigenpolarisations for both dielectric and homogeneous magnetic surfaces, we note how these could be utilised to find the total phase (related to the retardance of the surface) and amplitude of a reflected wave (with polarisation dependence of reflectance encompassed in the diattenuation of the surface) in conjunction with the eigenvalues for a surface \cite{gutierrez-vega_pancharatnamberry_2011, lu_homogeneous_1994}. For homogeneous magnetic surfaces, we highlight that the off-diagonal polarisation mixing terms introduce inhomogeneity, or non-orthogonality to the plane wave eigenpolarisations, and note how this can lead to an exceptional point where the eigenpolarisations become parallel at the same point the eigenvalues coalesce. We highlight the potential applications of utilising singularimetry to probe these EPs for both the plane wave and beam shift cases. 

This analysis is generally applicable to other surfaces, where for example the Jones matrices for any generic relativistically moving surface were recently derived in \cite{azar_fresnel_2025}. For both homogeneous and inhomogeneous Jones matrices, the inhomogeneity parameter and the full eigendecompsition are sufficient to fully characterise the reflected phase and amplitude \cite{lu_homogeneous_1994}. 

We utilise the magnetic Artmann operators we derived to find the eigenpolarisations for a unit order vortex beam reflecting from a homogeneous magnetic surface, finding spatially homogeneous polarisations for $l=\pm1$ vortex beams. We also note how the polarisation rearranges around the phase vortex upon reflection for non-eigenpolarisation inputs, and highlight the differences between the dielectric and magnetic polarisations. 
%The polarisation rearranging around the vortex was found to wrap around the entire polarisation sphere, akin to a 2D bimeron texture. Given the intensity near the vortex is very low, it is unlikely that this topological texture would be readily measurable. 

Limiting our expansion of the reflection matrix to first order in the Fourier beam spread parameter $\delta$, has allowed us to characterise both the total phase acquired and reflected amplitude of homogeneously polarised unit order vortex beams from homogeneous magnetic surfaces. This limitation however has restricted our treatment to only $l=\abs{1}$ scalar vortex beams. To treat higher order scalar vortex beams as done in \cite{dennis_topological_2012} would require expansion to higher order $\delta$ to probe.

Given the improvements to metrological methods achieved by OAM vortex beams in multiple disciplines \cite{pancaldi_high-resolution_2024}, and the potential for the magnetic plane wave EP to further enhance this, the natural extension would be to look at utilising singularimetry again as a method of surface metrology, in particular for the non-homogeneous, patterned surface magnetisations like those studied for the MHD effect \cite{fanciulli_magnetic_2022}. For extension to spatially non-homogeneous magnetic surfaces, these require expansion to orders $\delta > 1$, and there is some subtilty here in the orders of terms which must be kept for adequate approximation, which will be addressed in further work. It is also possible that there is further topological character and spatial inhomogeneity in the eigenpolarisations for higher order $l>\abs{1}$ scalar vortex beams, which would require extension to higher order $\delta$ to probe. 

This theory as presented allows for the eigenpolarisations of unit order scalar vortex beams to be calculated, revealing for which inputs the reflected beam will be homogeneously polarised. We also detail how the vortex centroid shift can be directly calculated from this theory, and highlight the differences between known dielectric shifts and those from homogeneous magnetic surfaces, and highlight the contribution of each individual magnetisation on these shifts. 

%when expanded to higher $\delta$ orders on the order of unknowns in the Jones matrix and reflecting an equal order $l$ vortex beam, would allow for surface magnetisation and other reflection characteristics to be measured directly from vortex positions as explored in \cite{dennis_topological_2012} for characterisation of dielectric surfaces. In particular, the ability to directly relate the positions of vortices after reflection to the aberration terms would prove useful for aberration correction \cite{zhai_turbulence_2020, zhang_machine_2019}, with the resulting higher order vortex beam eigenpolarisations providing aberration resistant polarisation states for these systems. The extension to non-homogeneous magnetic surfaces \cite{fanciulli_electromagnetic_2021}, and to $l>\abs{1}$ for these surfaces, will be detailed in further work. 

%\begin{itemize}
 %   \item \textcolor{Apricot}{Results section showing the magnetic eigenpolarisations of the full beam, plotted with the vortex centroid shift thing also plotted, alongside or superimposed on some post-selected/ non-eigenpolarisation ones? Also a graph here of how the plane wave eigenpolarisations of magnetic surface lead to a mixed polarisation beam. }
    
%\end{itemize}
\section{Acknowledgements}
MG acknowledges the University of Glasgow College of Science and Engineering Scholarship which supported this work. JBG acknowledges the hospitality of the Max Planck Institute for the Physics of Complex Systems, Dresden, and Leibniz Universität Hannover, whose support of an extended research visit enabled part of this work and helpful discussions with Mark Dennis and Marco Ornigotti. JBG wishes to acknowledge the support of the JSPS International Joint Research Program JRP- LEAD with UKRI under Grant No. JPJSJRP20241710.

\printbibliography
\newpage
\section{Supplementary: Lab to beam frame conversion}
The conversion between the lab frame coordinates $x_\sigma, y_\sigma$ and $z_\sigma$ and the beam frame coordinates $x,y$ and $z$ can be given by a simple rotation around the $y$-axis. The rotation matrix $\mathbf{Y}(\theta_0)$ which generates this rotation is given as:

  \begin{equation*}\label{rot_y}
    \mathbf{Y}(\theta_0) = 
    \begin{pmatrix}
    \cos{\theta_0} & 0 & \sin{\theta_0}  \\
    0 & 1 & 0\\
    -\sin{\theta_0}  & 0 & \cos{\theta_0} 
    \end{pmatrix}.
    \end{equation*}

The conversion of the Cartesian coordinates is therefore given as: 

\begin{equation*}
\begin{aligned}
x&=\cos{\theta_0}\, x_\sigma + \sin{\theta_0}\, z_\sigma,
 \\
 y&=y_\sigma,
 \\
 z&=-\sin{\theta_0}\, x_\sigma + \cos{\theta_0} \,z_\sigma.
\end{aligned}
\end{equation*}

In the sample/lab frame, the magnetisations can be decomposed from the original magnetisation vector $\mathbf{M}$ along the Cartesian directions as follows:

\begin{equation*}
\begin{gathered}
\mathbf{M}=
\begin{pmatrix}
  m_l\\m_t\\m_p  
\end{pmatrix}, \,\,\text{where}
 \\
 \mathbf{m_p}=m_p\,\mathbf{x_\sigma}, \qquad \mathbf{m_t}=m_t\,\mathbf{y_\sigma}, \qquad \mathbf{m_l}=m_l\,\mathbf{z_\sigma}.
\end{gathered}
\end{equation*}

Where these magnetisations can then be converted into the beam frame by rewriting the magnetisation vector in this basis: 

\begin{equation*}
\begin{aligned}
\mathbf{M_B}&=
\begin{pmatrix}
  m_l\cos{\theta_0} + m_p\sin{\theta_0}\\
  m_t\\
  -m_l\sin{\theta_0} + m_p\cos{\theta_0}  
\end{pmatrix},
 \\
 \mathbf{m_{p,B}}&=(m_l\cos{\theta_0} + m_p\sin{\theta_0})\,\mathbf{x}=m_{p,B}\,\mathbf{x}, 
 \\ \mathbf{m_{t,B}}&=m_t \,\mathbf{y},
 \\\mathbf{m_{l,B}}&=(-m_l\sin{\theta_0} + m_p\cos{\theta_0})\,\mathbf{z}=m_{l,B}\,\mathbf{z}.
\end{aligned}
\end{equation*}

In the text, the beam frame versions $m_{p,B}, m_{t,B}$ and $m_{l,B}$ of the magnetisation coefficients are quoted as $m_p, m_t$ and $m_l$. 
\newpage
\section{Supplementary: Full vortex centroid shift calculation}

Below is the calculation to find the full reflected field, Eq. \eqref{eigpol_original}. We consider only the $\alpha$ integral of Eq. \eqref{reflected dielectric field}, where we consider an input beam with OAM $l$, 

\begin{equation*}
    \begin{aligned}
      \psi(x,y)_R \,&=
      \int_0^{\pi/2}\int^\pi_{-\pi} \sigma(\delta) \delta \mathrm{e}^{\mathrm{i}kr\delta\cos(\alpha-\phi) -\mathrm{i}l\alpha} \,\mathbf{R}\mathbf{E} \,d\delta d\alpha,\\
      &=\int_0^{\pi/2} \sigma(\delta) \delta \underbrace{\int^\pi_{-\pi} \mathrm{e}^{\mathrm{i}kr\delta\cos(\alpha-\phi) -\mathrm{i}l\alpha} \,\mathbf{R}\mathbf{E}\,  d\alpha } d\delta,
    \end{aligned}
    \end{equation*}

where we substitute in Eq. \eqref{Artmann def},

 \begin{equation*}
 \begin{aligned}
      \mathbf{R}(\delta,\alpha) &= \mathbf{R}_0 + \delta\cos\alpha\,\mathbf{R}_\text{GH} + \delta\sin\alpha\,\mathbf{R}_\text{IF} + \mathcal{O}(\delta^2),\\
      &= \mathbf{R}_0  +  \delta\left( \frac{\mathrm{e}^{\mathrm{i} \alpha}+\mathrm{e}^{-\mathrm{i} \alpha}}{2}\right)\,\mathbf{R}_\text{GH} + \delta\left( \frac{\mathrm{e}^{\mathrm{i} \alpha}-\mathrm{e}^{-\mathrm{i} \alpha}}{2i}\right)\,\mathbf{R}_\text{IF},
    \end{aligned}
    \end{equation*}
now for an $l=1$ beam, evaluating only the $\alpha$ integral making the substitution $\alpha' \rightarrow \alpha-\phi$ ,

\begin{equation}\label{alpha intergal for cmn}
\begin{aligned}
 &\mathbf{R}_0\mathbf{E}\int^\pi_{-\pi} \text{\raisebox{.5pt}{\textcircled{\raisebox{-.9pt} {1}}}} \,d\alpha
 +\delta\,\frac{\mathbf{R}_\text{GH}}{2}\mathbf{E}\int^\pi_{-\pi} \text{\raisebox{.5pt}{\textcircled{\raisebox{-.9pt} {2}}}}\,d\alpha
 -\delta\,\frac{\mathrm{i}\mathbf{R}_\text{IF}}{2}\mathbf{E}\int^\pi_{-\pi} \text{\raisebox{.5pt}{\textcircled{\raisebox{-.9pt} {3}}}}\,d\alpha,
 \\
 &= \mathbf{R}_0\mathbf{E}\,\mathrm{e}^{-\mathrm{i}\phi}\int^{\pi-\phi}_{-\pi-\phi} \text{\raisebox{.5pt}{\textcircled{\raisebox{-.9pt} {4}}}}\, d\alpha'
 +\delta\,\frac{\mathbf{R}_\text{GH}}{2}\mathbf{E}\left(\,\mathrm{e}^{-\mathrm{i}0\phi}\int^{\pi-\phi}_{-\pi-\phi} \text{\raisebox{.5pt}{\textcircled{\raisebox{-.9pt} {5}}}}  \,d\alpha'  + \mathrm{e}^{-\mathrm{i}2\phi}\int^{\pi-\phi}_{-\pi-\phi} \text{\raisebox{.5pt}{\textcircled{\raisebox{-.9pt} {6}}}}\,  d\alpha'   \right)
 \\
  &-\delta\,\frac{\mathrm{i}\mathbf{R}_\text{IF}}{2}\mathbf{E}\left(\,\mathrm{e}^{-\mathrm{i}0\phi}\int^{\pi-\phi}_{-\pi-\phi} \text{\raisebox{.5pt}{\textcircled{\raisebox{-.9pt} {5}}}}\, d\alpha'  - \mathrm{e}^{-\mathrm{i}2\phi}\int^{\pi-\phi}_{-\pi-\phi} \text{\raisebox{.5pt}{\textcircled{\raisebox{-.9pt} {6}}}}\, d\alpha'   \right)
 \\
 &\text{where the integrands are},
 \\
 &\text{\raisebox{.5pt}{\textcircled{\raisebox{-.9pt} {1}}}}= \mathrm{e}^{\mathrm{i}kr\delta\cos(\alpha-\phi) -\mathrm{i}\alpha}
 \qquad \qquad \qquad \, \, \, 
 \text{\raisebox{.5pt}{\textcircled{\raisebox{-.9pt} {2}}}}= \mathrm{e}^{\mathrm{i}kr\delta\cos(\alpha-\phi) -\mathrm{i}\alpha} (\mathrm{e}^{\mathrm{i} \alpha}+\mathrm{e}^{-\mathrm{i} \alpha})
 \\
 &\text{\raisebox{.5pt}{\textcircled{\raisebox{-.9pt} {3}}}}=
 \mathrm{e}^{\mathrm{i}kr\delta\cos(\alpha-\phi) -\mathrm{i}\alpha} (\mathrm{e}^{\mathrm{i} \alpha}-\mathrm{e}^{-\mathrm{i} \alpha})
 \qquad
 \text{\raisebox{.5pt}{\textcircled{\raisebox{-.9pt} {4}}}}=\mathrm{e}^{\mathrm{i}kr\delta\cos(\alpha') -\mathrm{i}\alpha'}  
 \\
 &\text{\raisebox{.5pt}{\textcircled{\raisebox{-.9pt} {5}}}}=\mathrm{e}^{\mathrm{i}kr\delta\cos(\alpha') } 
 \qquad \qquad \qquad \qquad \, \, \,
 \text{\raisebox{.5pt}{\textcircled{\raisebox{-.9pt} {6}}}}=\mathrm{e}^{\mathrm{i}kr\delta\cos(\alpha')-\mathrm{i}2\alpha' } 
\end{aligned}
\end{equation}
where the integrals in the last line of Eq. \eqref{alpha intergal for cmn} can be identified as in \cite{dennis_analogy_2012} as a Bessel function of the first kind. The Bessel function definition given in general form for variable $x$ for positive $a$ orders

\begin{equation*}
\begin{gathered}
 J(x)_a=\frac{1}{2\pi \mathrm{i}^a}\int^\pi_{-\pi} \mathrm{e}^{\mathrm{i}x\cos(\alpha) +\mathrm{i}a\alpha}d\alpha.
\end{gathered}
\end{equation*}

Following from \cite{gotte_spin-orbit_2013}, identifying that for both positive and negative orders the integral can be replaced with: 

\begin{equation*}
\begin{gathered}
 \int^\pi_{-\pi} \mathrm{e}^{\mathrm{i}x\cos(\alpha) +\mathrm{i}a\alpha}d\alpha= J(x)_{\abs{a}} 2\pi \mathrm{i}^\abs{a},
\end{gathered}
\end{equation*}

where $a$ is now any positive integer. Making this substitution in Eq. \eqref{alpha intergal for cmn} gives

\begin{equation*}
\begin{gathered}
 2\pi \mathrm{i} \, \mathbf{R}_0\mathbf{E}\, \mathrm{e}^{-\mathrm{i}\phi} \,J(k\delta r)_{1}
 + \delta\pi\,\mathbf{R}_\text{GH}\mathbf{E}\left(J(k\delta r)_{0} - \mathrm{e}^{-\mathrm{i}2\phi}J(k\delta r)_{2}  \right)
 - \delta\pi \mathrm{i}\,\mathbf{R}_\text{IF}\mathbf{E}\left(J(k\delta r)_{0} + \mathrm{e}^{-\mathrm{i}2\phi}J(k\delta r)_{2}   \right).
\end{gathered}
\end{equation*}

To further simplify this expression, $\delta \ll 1$ allows us to approximate the Bessel function close to the axis, making use of the sum representation of the function

\begin{equation*}
\begin{gathered}
 J(x)_a= \sum_{b=0}^\infty\frac{(-1)^b}{ 2^{2b+a}(a+b)!(b!)}x^{2b+a},
\end{gathered}
\end{equation*}
we can take only the term where $b=0$ as any higher terms are in the next order of $\delta$, and hence significantly smaller due to the $\delta$ term in the argument $x$. As a result, an approximation of the Bessel function valid for small $\delta$ is

\begin{equation*}
\begin{gathered}
 J(x)_a=\frac{1}{ 2^{a}(a)!}x^{a}.
\end{gathered}
\end{equation*}

Evaluating for terms up to $J(k\delta r)_2$ returns 

\begin{equation*}
\begin{gathered}
\pi \mathrm{i} \, \mathbf{R}_0\mathbf{E}\,\mathrm{e}^{-\mathrm{i}\phi}(k\delta r)
 + \delta \pi\,\mathbf{R}_\text{GH}\mathbf{E}\left( 1 -\mathrm{e}^{-\mathrm{i}2\phi} \frac{(k \delta r)^2}{8}  \right)
 - \delta \pi \mathrm{i} \,\mathbf{R}_\text{IF}\mathbf{E}\left( 1 + \mathrm{e}^{-\mathrm{i}2\phi} \frac{(k \delta r)^2}{8} \right).
\end{gathered}
\end{equation*}

Keeping only terms up to $\mathcal{O}(\delta)$ and substituting in $\eta^a=(kr)^a\mathrm{e}^{-\mathrm{i}a\phi}$ for $a \ge0$ or $\eta^{*a}=(kr)^a\mathrm{e}^{+\mathrm{i}a\phi}$ for $a < 0$ produces Eq. \eqref{eigpol_original}

\begin{equation*}
\begin{aligned}
  \psi(x,y)_R \, &\propto \, \mathrm{i} \delta \eta \,\mathbf{R}_0\mathbf{E} + \delta\,\mathbf{R}_\text{GH}\mathbf{E} -
  \mathrm{i}\delta \,\mathbf{R}_\text{IF}\mathbf{E},
  \\
  &=\delta \left(\mathrm{i}\eta \mathbf{R}_0 +\left( \mathbf{R}_\text{GH}-\mathrm{i}\mathbf{R}_\text{IF}\right)\right)\mathbf{E}, \,l \ge0,
  \\
  &=\delta \left(\mathrm{i}\eta^* \mathbf{R}_0 +\left( \mathbf{R}_\text{GH}+\mathrm{i}\mathbf{R}_\text{IF}\right)\right)\mathbf{E}, \,l < 0.
\end{aligned}
\end{equation*}

\newpage

\end{document}